\documentclass[%
 prd, twocolumn,
superscriptaddress,
nofootinbib,
 amsmath,amssymb,
 aps,
]{revtex4-2}

\usepackage{booktabs}
\usepackage{graphicx}
\usepackage{dcolumn}
\usepackage{bm}
\usepackage[colorlinks=true]{hyperref}
\usepackage{multirow}
\usepackage{xcolor}

\usepackage{subfigure}

{}

{}
{}

\newcommand{\aap}{{A\&A}}

\newcommand{\apjl}{{ApJL}}
\newcommand{\apjs}{{ApJS}}
\newcommand{\mnras}{{MNRAS}}

\usepackage{aas_macros}
\usepackage{orcidlink} 

\begin{document}


\title{Searches for signatures of ultra-light axion dark matter in polarimetry data of the European Pulsar Timing Array}

\author{N.~K.~Porayko\orcidlink{0000-0002-6955-8040}}
\email{nataliya.porayko@unimib.it}
\affiliation{Dipartimento di Fisica ``G. Occhialini", Universit{\'a} degli Studi di Milano-Bicocca, Piazza della Scienza 3, I-20126 Milano, Italy}
\affiliation{Max-Planck-Institut f{\"u}r Radioastronomie, Auf dem H{\"u}gel 69, 53121 Bonn, Germany} 
\author{P.~Usynina}%
\email{usynina.pg19@physics.msu.ru}
\affiliation{%
Sternberg Astronomical Institute, Moscow State University, Universitetsky pr., 13, Moscow 119234, Russia\\
}%
\author{J.~Terol-Calvo
\orcidlink{0000-0003-3117-5017}}%
\email{jortecal@protonmail.com}
\affiliation{%
 Instituto de Astrof\'isica de Canarias, C/ V\'ia L\'actea, s/n
E38205 - La Laguna, Tenerife, Spain
}%
\affiliation{%
Universidad de La Laguna, Dpto. Astrof\'isica, E38206 - La Laguna, Tenerife, Spain
}%
\affiliation{%
Istituto Nazionale di Fisica Nucleare, Sezione di Torino, via P. Giuria 1, I–10125 Torino, Italy
}%
\author{J.~Martin~Camalich
\orcidlink{https://orcid.org/0000-0002-5673-4500}}%
\affiliation{%
 Instituto de Astrof\'isica de Canarias, C/ V\'ia L\'actea, s/n
E38205 - La Laguna, Tenerife, Spain
}%
\affiliation{%
Universidad de La Laguna, Dpto. Astrof\'isica, E38206 - La Laguna, Tenerife, Spain
}%
\author{G.~M.~Shaifullah\orcidlink{0000-0002-8452-4834}}
\affiliation{Dipartimento di Fisica ``G. Occhialini", Universit{\'a} degli Studi di Milano-Bicocca, Piazza della Scienza 3, I-20126 Milano, Italy}
\affiliation{INFN, Sezione di Milano-Bicocca, Piazza della Scienza 3, I-20126 Milano, Italy}
\affiliation{INAF - Osservatorio Astronomico di Cagliari, via della Scienza 5, 09047 Selargius (CA), Italy}
\author{A. Castillo\orcidlink{0000-0002-9887-3293}}%
\affiliation{
 Departamento de F\'isica, Universidad Antonio Nari\~no.
 Cra 3 Este \# 45-03, Bogot\'a, Colombia.}%
\author{D. Blas\orcidlink{0000-0003-2646-0112}}%
\affiliation{Institut de F\'isica d’Altes Energies (IFAE), The Barcelona Institute of Science and Technology, Campus UAB, 08193 Bellaterra (Barcelona), Spain}
\affiliation{Instituci\'o Catalana de Recerca i Estudis Avan\c cats (ICREA), Passeig Llu\'is Companys 23, 08010 Barcelona, Spain}
\author{L.~Guillemot\orcidlink{0000-0002-9049-8716}}
\affiliation{LPC2E, OSUC, Univ Orleans, CNRS, CNES, Observatoire de Paris, F-45071 Orleans, France}
\affiliation{Observatoire Radioastronomique de Nan\c{c}ay, Observatoire de Paris, Universit\'e PSL, Université d'Orl\'eans, CNRS, 18330 Nan\c{c}ay, France}
\author{M. Peel\orcidlink{0000-0003-3412-2586}}%
\affiliation{%
 Imperial College London, Blackett Lab, Prince Consort Road, London SW7 2AZ, UK
}%
\author{C.~Tiburzi\orcidlink{0000-0001-6651-4811}}
\affiliation{INAF - Osservatorio Astronomico di Cagliari, via della Scienza 5, 09047 Selargius (CA), Italy}
\author{K. Postnov\orcidlink{0000-0002-1705-617X}}%
\affiliation{%
Sternberg Astronomical Institute, Moscow State University, Universitetsky pr., 13, Moscow 119234, Russia\\
}%
\affiliation{%
Institute for Nuclear Research,  Moscow 117312, Russia
}%
\author{M.~Kramer\orcidlink{0000-0002-4175-2271}}
\affiliation{Max-Planck-Institut f{\"u}r Radioastronomie, Auf dem H{\"u}gel 69, 53121 Bonn, Germany}
\affiliation{Jodrell Bank Centre for Astrophysics, Department of Physics and Astronomy, University of Manchester, Manchester M13 9PL, UK}
\author{J.~Antoniadis\orcidlink{0000-0003-4453-776}}
\affiliation{Institute of Astrophysics, FORTH, N. Plastira 100, 70013, Heraklion, Greece} 
\affiliation{Max-Planck-Institut f{\"u}r Radioastronomie, Auf dem H{\"u}gel 69, 53121 Bonn, Germany}

\author{S.~Babak\orcidlink{0000-0001-7469-4250}}
\affiliation{Universit{\'e} Paris Cit{\'e} CNRS, Astroparticule et Cosmologie, 75013 Paris, France}

\author{A.-S.~Bak~Nielsen\orcidlink{ 0000-0002-1298-9392}}
\affiliation{Max-Planck-Institut f{\"u}r Radioastronomie, Auf dem H{\"u}gel 69, 53121 Bonn, Germany}
\affiliation{Fakult{\"a}t f{\"u}r Physik, Universit{\"a}t Bielefeld, Postfach 100131, 33501 Bielefeld, Germany}

\author{E.~Barausse\orcidlink{0000-0001-6499-6263 }}
\affiliation{SISSA — International School for Advanced Studies, Via Bonomea 265, 34136, Trieste, Italy and INFN, Sezione di Trieste}
\affiliation{IFPU — Institute for Fundamental Physics of the Universe, Via Beirut 2, 34014 Trieste, Italy}

\author{C.~G.~Bassa\orcidlink{0000-0002-1429-9010}}
\affiliation{ASTRON, Netherlands Institute for Radio Astronomy, Oude Hoogeveensedijk 4, 7991 PD, Dwingeloo, The Netherlands}

\author{C.~Blanchard}
\affiliation{LPC2E, OSUC, Univ Orleans, CNRS, CNES, Observatoire de Paris, F-45071 Orleans, France}
\affiliation{Observatoire Radioastronomique de Nan\c{c}ay, Observatoire de Paris, Universit\'e PSL, Université d'Orl\'eans, CNRS, 18330 Nan\c{c}ay, France}

\author{M.~Bonetti\orcidlink{0000-0001-7889-6810}}
\affiliation{Dipartimento di Fisica ``G. Occhialini", Universit{\'a} degli Studi di Milano-Bicocca, Piazza della Scienza 3, I-20126 Milano, Italy}
\affiliation{INFN, Sezione di Milano-Bicocca, Piazza della Scienza 3, I-20126 Milano, Italy}
\affiliation{INAF - Osservatorio Astronomico di Brera, via Brera 20, I-20121 Milano, Italy}

\author{E.~Bortolas\orcidlink{0000-0001-9458-821X}}
\affiliation{Dipartimento di Fisica ``G. Occhialini", Universit{\'a} degli Studi di Milano-Bicocca, Piazza della Scienza 3, I-20126 Milano, Italy}
\affiliation{INFN, Sezione di Milano-Bicocca, Piazza della Scienza 3, I-20126 Milano, Italy}
\affiliation{INAF - Osservatorio Astronomico di Padova,  Vicolo dell'Osservatorio, 5, I-35122 Padova (PD), Italy}

\author{P.~R.~Brook\orcidlink{0000-0003-3053-6538}}
\affiliation{Institute for Gravitational Wave Astronomy and School of Physics and Astronomy, University of Birmingham, Edgbaston, Birmingham B15 2TT, UK}

\author{M.~Burgay\orcidlink{0000-0002-8265-4344}}
\affiliation{INAF - Osservatorio Astronomico di Cagliari, via della Scienza 5, 09047 Selargius (CA), Italy}

\author{R.~N.~Caballero\orcidlink{0000-0001-9084-9427}}
\affiliation{Hellenic Open University, School of Science and Technology, 26335 Patras, Greece}

\author{A.~Chalumeau\orcidlink{0000-0003-2111-1001}}
\affiliation{ASTRON, Netherlands Institute for Radio Astronomy, Oude Hoogeveensedijk 4, 7991 PD, Dwingeloo, The Netherlands}

\author{D.~J.~Champion\orcidlink{0000-0003-1361-7723}}
\affiliation{Max-Planck-Institut f{\"u}r Radioastronomie, Auf dem H{\"u}gel 69, 53121 Bonn, Germany}

\author{S.~Chanlaridis\orcidlink{0000-0002-9323-9728}}
\affiliation{Institute of Astrophysics, FORTH, N. Plastira 100, 70013, Heraklion, Greece} 

\author{S.~Chen\orcidlink{0000-0002-3118-5963}}
\affiliation{Shanghai Astronomical Observatory, Chinese Academy of Sciences, Shanghai 200030, P.~R.~China}

\author{I.~Cognard\orcidlink{0000-0002-1775-9692}}
\affiliation{LPC2E, OSUC, Univ Orleans, CNRS, CNES, Observatoire de Paris, F-45071 Orleans, France}
\affiliation{Observatoire Radioastronomique de Nan\c{c}ay, Observatoire de Paris, Universit\'e PSL, Université d'Orl\'eans, CNRS, 18330 Nan\c{c}ay, France}

\author{G.~Desvignes\orcidlink{0000-0003-3922-4055}}
\affiliation{Max-Planck-Institut f{\"u}r Radioastronomie, Auf dem H{\"u}gel 69, 53121 Bonn, Germany}

\author{M.~Falxa}
\affiliation{LPC2E, OSUC, Univ Orleans, CNRS, CNES, Observatoire de Paris, F-45071 Orleans, France}
\affiliation{Universit{\'e} Paris Cit{\'e} CNRS, Astroparticule et Cosmologie, 75013 Paris, France}

\author{R.~D.~Ferdman}
\affiliation{School of Physics, Faculty of Science, University of East Anglia, Norwich NR4 7TJ, UK}

\author{A.~Franchini\orcidlink{0000-0002-8400-0969}}
\affiliation{Dipartimento di Fisica ``G. Occhialini", Universit{\'a} degli Studi di Milano-Bicocca, Piazza della Scienza 3, I-20126 Milano, Italy}
\affiliation{INFN, Sezione di Milano-Bicocca, Piazza della Scienza 3, I-20126 Milano, Italy}

\author{J.~R.~Gair\orcidlink{0000-0002-1671-3668}}
\affiliation{Max Planck Institute for Gravitational Physics (Albert Einstein Institute), Am Mu{\"u}hlenberg 1, 14476 Potsdam, Germany}

\author{A.~Golden\orcidlink{0000-0001-8208-4292}}
\affiliation{Ollscoil na Gaillimhe --- University of Galway, University Road, Galway, H91 TK33, Ireland}

\author{B. Goncharov\orcidlink{0000-0003-3189-5807}}%
\affiliation{Gran Sasso Science Institute (GSSI), I-67100 L'Aquila, Italy}
\affiliation{INFN, Laboratori Nazionali del Gran Sasso, I-67100 Assergi, Italy}

\author{E.~Graikou}
\affiliation{Max-Planck-Institut f{\"u}r Radioastronomie, Auf dem H{\"u}gel 69, 53121 Bonn, Germany}

\author{J.-M.~Grie{\ss}meier\orcidlink{0000-0003-3362-7996}}
\affiliation{LPC2E, OSUC, Univ Orleans, CNRS, CNES, Observatoire de Paris, F-45071 Orleans, France}
\affiliation{Observatoire Radioastronomique de Nan\c{c}ay, Observatoire de Paris, Universit\'e PSL, Université d'Orl\'eans, CNRS, 18330 Nan\c{c}ay, France}


\author{Y.~J.~Guo}
\affiliation{Max-Planck-Institut f{\"u}r Radioastronomie, Auf dem H{\"u}gel 69, 53121 Bonn, Germany}

\author{H.~Hu\orcidlink{0000-0002-3407-8071}}
\affiliation{Max-Planck-Institut f{\"u}r Radioastronomie, Auf dem H{\"u}gel 69, 53121 Bonn, Germany}

\author{F.~Iraci}
\affiliation{INAF - Osservatorio Astronomico di Cagliari, via della Scienza 5, 09047 Selargius (CA), Italy}
\affiliation{Universit{\'a} di Cagliari, Dipartimento di Fisica, S.P. Monserrato-Sestu Km 0,700 - 09042 Monserrato (CA), Italy}

\author{D.~Izquierdo-Villalba\orcidlink{0000-0002-6143-1491}}
\affiliation{Dipartimento di Fisica ``G. Occhialini", Universit{\'a} degli Studi di Milano-Bicocca, Piazza della Scienza 3, I-20126 Milano, Italy}
\affiliation{INFN, Sezione di Milano-Bicocca, Piazza della Scienza 3, I-20126 Milano, Italy}

\author{J.~Jang\orcidlink{0000-0003-4454-0204}}
\affiliation{Max-Planck-Institut f{\"u}r Radioastronomie, Auf dem H{\"u}gel 69, 53121 Bonn, Germany}

\author{J.~Jawor\orcidlink{0000-0003-3391-0011}}
\affiliation{Max-Planck-Institut f{\"u}r Radioastronomie, Auf dem H{\"u}gel 69, 53121 Bonn, Germany}

\author{G.~H.~Janssen\orcidlink{0000-0003-3068-3677}}
\affiliation{ASTRON, Netherlands Institute for Radio Astronomy, Oude Hoogeveensedijk 4, 7991 PD, Dwingeloo, The Netherlands}
\affiliation{Department of Astrophysics/IMAPP, Radboud University Nijmegen, P.O. Box 9010, 6500 GL Nijmegen, The Netherlands}

\author{A.~Jessner}
\affiliation{Max-Planck-Institut f{\"u}r Radioastronomie, Auf dem H{\"u}gel 69, 53121 Bonn, Germany}

\author{R.~Karuppusamy\orcidlink{0000-0002-5307-2919}}
\affiliation{Max-Planck-Institut f{\"u}r Radioastronomie, Auf dem H{\"u}gel 69, 53121 Bonn, Germany}

\author{E.~F.~Keane\orcidlink{0000-0002-4553-655X}}
\affiliation{School of Physics, Trinity College Dublin, College Green, Dublin 2, D02 PN40, Ireland}

\author{M.~J.~Keith\orcidlink{0000-0001-5567-5492}}
\affiliation{Jodrell Bank Centre for Astrophysics, Department of Physics and Astronomy, University of Manchester, Manchester M13 9PL, UK}


\author{M.~A.~Krishnakumar\orcidlink{0000-0003-4528-2745}}
\affiliation{Max-Planck-Institut f{\"u}r Radioastronomie, Auf dem H{\"u}gel 69, 53121 Bonn, Germany}
\affiliation{Fakult{\"a}t f{\"u}r Physik, Universit{\"a}t Bielefeld, Postfach 100131, 33501 Bielefeld, Germany}

\author{K.~Lackeos\orcidlink{0000-0002-6554-3722}}
\affiliation{Max-Planck-Institut f{\"u}r Radioastronomie, Auf dem H{\"u}gel 69, 53121 Bonn, Germany}

\author{K.~J.~Lee}
\affiliation{Department of Astronomy, School of Physics, Peking University, Beijing 100871, P.~R.~China} 
\affiliation{National Astronomical Observatories, Chinese Academy of Sciences, Beijing 100101, P.~R.~China}
\affiliation{Yunnan Astronomical Observatories, Chinese Academy of Sciences, Kunming 650216, Yunnan, P.~R.~China}
\affiliation{Beijing Laser Acceleration Innovation Center, Huairou, Beijing 101400, P.~R.~China}

\author{K.~Liu\orcidlink{0000-0003-4453-776}}
\affiliation{Shanghai Astronomical Observatory, Chinese Academy of Sciences, Shanghai 200030, P.~R.~China}
\affiliation{Max-Planck-Institut f{\"u}r Radioastronomie, Auf dem H{\"u}gel 69, 53121 Bonn, Germany}

\author{Y.~Liu\orcidlink{0000-0001-9986-9360}}
\affiliation{Fakult{\"a}t f{\"u}r Physik, Universit{\"a}t Bielefeld, Postfach 100131, 33501 Bielefeld, Germany}
\affiliation{National Astronomical Observatories, Chinese Academy of Sciences, Beijing 100101, P.~R.~China}

\author{A.~G.~Lyne}
\affiliation{Jodrell Bank Centre for Astrophysics, Department of Physics and Astronomy, University of Manchester, Manchester M13 9PL, UK}

\author{J.~W.~McKee\orcidlink{0000-0002-2885-8485}}
\affiliation{E. A. Milne Centre for Astrophysics, University of Hull, Cottingham Road, Kingston-upon-Hull, HU6 7RX, UK}
\affiliation{Centre of Excellence for Data Science, Artificial Intelligence and Modelling (DAIM), University of Hull, Cottingham Road, Kingston-upon-Hull, HU6 7RX, UK}

\author{R.~A.~Main}
\affiliation{Max-Planck-Institut f{\"u}r Radioastronomie, Auf dem H{\"u}gel 69, 53121 Bonn, Germany}

\author{M.~B.~Mickaliger\orcidlink{0000-0001-6798-5682}}
\affiliation{Jodrell Bank Centre for Astrophysics, Department of Physics and Astronomy, University of Manchester, Manchester M13 9PL, UK}

\author{I.~C.~Ni\c{t}u\orcidlink{0000-0003-3611-3464}}
\affiliation{Jodrell Bank Centre for Astrophysics, Department of Physics and Astronomy, University of Manchester, Manchester M13 9PL, UK}

\author{A.~Parthasarathy\orcidlink{0000-0002-4140-5616}}
\affiliation{Max-Planck-Institut f{\"u}r Radioastronomie, Auf dem H{\"u}gel 69, 53121 Bonn, Germany}

\author{B.~B.~P.~Perera\orcidlink{0000-0002-8509-5947}}
\affiliation{Arecibo Observatory, HC3 Box 53995, Arecibo, PR 00612, USA}

\author{D.~Perrodin\orcidlink{0000-0002-1806-2483}}
\affiliation{INAF - Osservatorio Astronomico di Cagliari, via della Scienza 5, 09047 Selargius (CA), Italy}

\author{A.~Petiteau\orcidlink{0000-0002-7371-9695}}
\affiliation{IRFU, CEA, Université Paris-Saclay, F-91191 Gif-sur-Yvette, France}
\affiliation{Universit{\'e} Paris Cit{\'e} CNRS, Astroparticule et Cosmologie, 75013 Paris, France}


\author{A.~Possenti}
\affiliation{INAF - Osservatorio Astronomico di Cagliari, via della Scienza 5, 09047 Selargius (CA), Italy}

\author{H.~Quelquejay~Leclere\orcidlink{0000-0002-6766-2004}}
\affiliation{Universit{\'e} Paris Cit{\'e} CNRS, Astroparticule et Cosmologie, 75013 Paris, France}

\author{A.~Samajdar\orcidlink{0000-0002-0857-6018}}
\affiliation{Institut f\"{u}r Physik und Astronomie, Universit\"{a}t Potsdam, Haus 28, Karl-Liebknecht-Str. 24/25, 14476, Potsdam, Germany}

\author{S.~A.~Sanidas}
\affiliation{Jodrell Bank Centre for Astrophysics, Department of Physics and Astronomy, University of Manchester, Manchester M13 9PL, UK}

\author{A.~Sesana}
\affiliation{Dipartimento di Fisica ``G. Occhialini", Universit{\'a} degli Studi di Milano-Bicocca, Piazza della Scienza 3, I-20126 Milano, Italy}
\affiliation{INFN, Sezione di Milano-Bicocca, Piazza della Scienza 3, I-20126 Milano, Italy}
\affiliation{INAF - Osservatorio Astronomico di Brera, via Brera 20, I-20121 Milano, Italy}


\author{L.~Speri\orcidlink{0000-0002-5442-7267}}
\affiliation{Max Planck Institute for Gravitational Physics (Albert Einstein Institute), Am Mu{\"u}hlenberg 1, 14476 Potsdam, Germany}

\author{R.~Spiewak}
\affiliation{Jodrell Bank Centre for Astrophysics, Department of Physics and Astronomy, University of Manchester, Manchester M13 9PL, UK}

\author{B.~W.~Stappers}
\affiliation{Jodrell Bank Centre for Astrophysics, Department of Physics and Astronomy, University of Manchester, Manchester M13 9PL, UK}

\author{S.~C.~Susarla\orcidlink{0000-0003-4332-8201}}
\affiliation{Ollscoil na Gaillimhe --- University of Galway, University Road, Galway, H91 TK33, Ireland}

\author{G.~Theureau\orcidlink{0000-0002-3649-276X}}
\affiliation{LPC2E, OSUC, Univ Orleans, CNRS, CNES, Observatoire de Paris, F-45071 Orleans, France}
\affiliation{Observatoire Radioastronomique de Nan\c{c}ay, Observatoire de Paris, Universit\'e PSL, Université d'Orl\'eans, CNRS, 18330 Nan\c{c}ay, France}
\affiliation{Laboratoire Univers et Th{\'e}ories LUTh, Observatoire de Paris, Universit{\'e} PSL, CNRS, Universit{\'e} de Paris, 92190 Meudon, France}


\author{E.~van~der~Wateren\orcidlink{0000-0003-0382-8463}}
\affiliation{Department of Astrophysics/IMAPP, Radboud University Nijmegen, P.O. Box 9010, 6500 GL Nijmegen, The Netherlands}
\affiliation{ASTRON, Netherlands Institute for Radio Astronomy, Oude Hoogeveensedijk 4, 7991 PD, Dwingeloo, The Netherlands}

\author{A.~Vecchio\orcidlink{0000-0002-6254-1617}}
\affiliation{Institute for Gravitational Wave Astronomy and School of Physics and Astronomy, University of Birmingham, Edgbaston, Birmingham B15 2TT, UK}

\author{V.~Venkatraman~Krishnan\orcidlink{0000-0001-9518-9819}}
\affiliation{Max-Planck-Institut f{\"u}r Radioastronomie, Auf dem H{\"u}gel 69, 53121 Bonn, Germany}

\author{J.~Wang\orcidlink{0000-0003-1933-6498}}
\affiliation{Ruhr University Bochum, Faculty of Physics and Astronomy, Astronomical Institute (AIRUB), 44780 Bochum, Germany}

\author{L.~Wang}
\affiliation{Jodrell Bank Centre for Astrophysics, Department of Physics and Astronomy, University of Manchester, Manchester M13 9PL, UK}

\author{Z.~Wu\orcidlink{0000-0002-1381-7859}}
\affiliation{National Astronomical Observatories, Chinese Academy of Sciences, Beijing 100101, P.~R.~China}

\collaboration{EPTA Collaboration}\noaffiliation

\date{\today}

\begin{abstract}
Ultra-light axion-like particles (ALPs) can be a viable solution to the dark matter problem. The scalar field associated with ALPs, coupled to the electromagnetic field, acts as an active birefringent medium, altering the polarisation properties of light through which it propagates. In particular, oscillations of the axionic field induce monochromatic variations of the plane of linearly polarised radiation of astrophysical signals. The radio emission of millisecond pulsars provides an excellent tool to search for such manifestations, given their high fractional linear polarisation and negligible fluctuations of their polarisation properties. We have searched for the evidence of ALPs in the polarimetry measurements of pulsars collected and preprocessed for the European Pulsar Timing Array (EPTA) campaign. Focusing on the twelve brightest sources in linear polarisation, we searched for an astrophysical signal from axions using both frequentist and Bayesian statistical frameworks. For the frequentist analysis, which uses Lomb-Scargle periodograms at its core, no statistically significant signal has been found. The model used for the Bayesian analysis has been adjusted to accommodate multiple deterministic systematics that may be present in the data. A statistically significant signal has been found in the dataset of multiple pulsars with common frequency between $10^{-8}$~Hz and $2\times10^{-8}$~Hz, which can most likely be explained by the residual Faraday rotation in the terrestrial ionosphere. Strong bounds on the coupling constant $g_{a\gamma}$, in the same ballpark as other searches, have been obtained in the mass range between $6\times10^{-24}$~eV and $5\times10^{-21}$~eV. We conclude by discussing problems that can limit the sensitivity of our search for ultra-light axions in the polarimetry data of pulsars, and possible ways to resolve them.
\end{abstract}

\maketitle

\section{Introduction}

The search for dark matter (DM) continues to be among the most active research fields in fundamental 
physics, astrophysics and cosmology \cite{2018RvMP...90d5002B,2021PrPNP.11903865A}. In recent years, different models have been scrutinized and their parameter space constrained by a combination of novel phenomenological approaches and improvements in the quantity and quality of data. Regarding the first aspect, weakly interacting particles with very light masses (of about $10^{-22}$ eV),  which can be non-thermal cosmological relics, are playing a special role as candidates for ultra-light dark matter (ULDM) \cite{1983PhLB..120..127P,
1983PhLB..120..133A,1983PhLB..120..137D,
2017PhRvD..95d3541H}. Irrespective of the DM problem, such pseudo-scalar Goldstone bosons have a possible connection to the strong-CP problem in QCD (the QCD axion) and appear in various Standard Model extensions of particle physics \cite{2010ARNPS..60..405J}. They have a distinct impact on astrophysical observables as compared to models of DM particles with masses above $\sim 1$\,eV \cite{MARSH20161}. Very low masses of ULDM particles imply that their de-Broglie wave is much larger than their particle separation, which enables a purely classical description (sometimes referred to as `wave DM', see \cite{2021ARA&A..59..247H} for a review and references).  As far as data is concerned, here we will focus on astrophysical observations with an ever-growing capacity to study new phenomena with increasingly high accuracy. Precise measurements of pulsars play a privileged role in this program. 

A remarkable experiment probing ultra-light axion-like particles (ALPs) using pulsars is based on the search for periodic changes in the time-of-arrivals (TOAs) of pulses propagating through the axionic field. The condensate of the ULDM particles with phenomenologically interesting masses $m_a\sim 10^{-23}-10^{-22}$~eV, forms clumps with a typical size of the de-Broglie wavelength of $\sim60~\textrm{pc}\times(10^{-22}\textrm{eV}/m_a)$, where $m_a$ is the axionic mass. Within each clump, ULDM experiences fast coherent oscillation with a typical frequency defined by $m_a$, inducing variations of the space-time metric with twice the frequency. This effect was originally described by \cite{2014JCAP...02..019K} and was searched for in the timing data of various Pulsar Timing Arrays (PTAs) \cite{2014PhRvD..90f2008P, 2018PhRvD..98j2002P, 2020JCAP...09..036K, 2020EPJC...80..419N, 2022PhRvR...4a2022X, 2023ApJ...951L..11A}. The tightest bounds to date using this method have been obtained in~\cite{2023PhRvL.131q1001S}, constraining the ULDM density to be below a few tenths of the observed DM abundance in the mass range $m_a\sim[10^{-24}~\textrm{eV}, 10^{-23.3}~\textrm{eV}]$. 

If ULDM couples non-gravitationally to Standard Model particles, it will affect the spin of pulsars as well as the frequency of the atomic clocks used to record pulsar timing measurements, leading to another monochromatic signal in the timing observations at the frequency $\omega=m_a$ \cite{2016PhRvD..93g5029G, 2022PhRvD.106c5032K, 2023ApJ...951L..11A,Smarra:2024kvv}. In addition to coherent oscillations, ULDM spatial density fluctuations in the Galactic halo source stochastic perturbations of the TOAs at frequencies $\omega\lesssim m_a\sigma^2$  enabling searches for ALPs in the mass range $m_a\sim[10^{-18}~\textrm{eV},10^{-14}~\textrm{eV}]$~\cite{Smarra:2024kvv}.

In the present work, we go beyond the aforementioned searches for ULDM in the timing residuals of PTAs and look for the complementary effect of ALPs emerging in the polarimetry data. In particular, the \textit{effective} interaction between photons and ALPs, characterised by the dimensionful coupling $g_{a\gamma}$,\footnote{
The photon-ALP interaction in Eq.~\eqref{eq:Lagrangian} corresponds to a nonrenormalizable operator with a coupling $g_{a\gamma}$ treated within the framework of effective field theories. In this context, the interaction represents the low-energy imprint of new interactions and degrees of freedom that become dynamical at energy scales roughly $\gtrsim 1/g_{a\gamma}$ (see Ref.~\cite{Castillo:2022zfl} for further discussion and caveats).
} induces sinusoidal oscillations at frequency $\omega=m_a$ of the polarisation plane of linearly polarised emission of astrophysical signals. This effect, also known as \textit{cosmic birefringence} \cite{1990PhRvD..41.1231C, 1991PhRvD..43.3789C}, allows to impose constraint on $g_{a\gamma}$ through multiple astrophysical probes, including polarisation measurements of parsec-scale jets in active galaxies~\cite{Ivanov:2018byi}, searches for birefringence effect in the polarisation of the cosmic microwave background (CMB)~\cite{BICEPKeck:2021sbt}, the non-observation of a reduction of the net polarised fraction of the CMB light as compared to the $\Lambda$CDM prediction~\cite{Fedderke:2019ajk}. In addition, two other types of experiments that examine axion production in extreme environments, i.e. solar axion searches~\cite{Cast2017} and searching for ALPs-to-photon conversion in the supernova SN~1987A~\cite{Payez:2014xsa}, enable to set constraints on the coupling constant which are independent of the relic axion abundance (local DM density), in contrast to the other probes mentioned above. Pulsars, which emit highly linearly polarised radiation with unprecedented temporal stability, offer an exciting opportunity to probe and directly detect cosmic birefringence due to ultra-light axions \cite{2020PhRvD.101f3012L}. The latest limits from pulsar polarimetry were obtained using the data collected within the Parkes PTA campaign \cite{csf+2019, Castillo:2022zfl}. Here, based on previous studies, for the first time we investigate the presence of ULDM signatures in the polarisation data of the European PTA (EPTA) covering a total period of 11 years. To this aim, we analyse the polarisation calibrated EPTA dataset, clean it from known systematics, i.e. from the effects of the terrestrial plasma, and construct the time series of position angles (PAs), which characterise the orientation of the plane of linearly polarised pulsar radiation. Finally, we search for ULDM signals in the PA time series and set upper limits on the coupling constant applying both frequentist and Bayesian data analysis methods.

The paper is organised as follows. In Sec.~\ref{sec:birefringence}, we introduce the concept of ULDM and its effect on pulsar polarisation data. In Sec.~\ref{sec:obs_epta}, we introduce the EPTA dataset and discuss the essential steps of data curation. Sec.~\ref{sec:method} is devoted to the description of the techniques developed to analyse the data, separately by frequentist and Bayesian methods. Our main results are presented in Sec.~\ref{sec:result}. Finally, we summarise our study in Sec.~\ref{sec:conclusion}. Throughout the paper, we work in the natural system of units $\hbar=c=1$.

\section{Birefringence of pulsar light due to ultra-light axions} \label{sec:birefringence}

A medium formed by an ALP relic\footnote{Meaning that axions are produced in the early Universe through various mechanism, see, e.g. \cite{2023JCAP...05..028A}.} can affect the propagation of electromagnetic (EM) waves by rotating their polarisation plane~\cite{Carroll:1989vb,Harari:1992ea}. This effect arises from the Lagrangian describing the ALP-photon interactions,
\begin{equation}\label{eq:Lagrangian}
    \mathcal{L}= -\frac{1}{4}F_{\mu\nu}F^{\mu\nu}+\frac{g_{a\gamma}}{4} aF_{\mu\nu}\tilde{F}^{\mu\nu}+\frac{1}{2}\left(\partial_{\mu}a\partial^{\mu}a-m_{a}^{2} a^2\right)\,,
\end{equation}
where $a$ is the ALP field, $F_{\mu\nu}$ is the EM Maxwell tensor and $\Tilde{F}^{\mu\nu}$ is its dual. Note that the coupling constant $g_{a\gamma}$ has units of inverse mass. 

The solution to the modified Maxwell equations in the adiabatic (WKB) approximation\footnote{Assuming that variations of the ALP field are much longer than the ALP oscillation period $T_a=2\pi/m_a\sim 4\times 10^7\, \mathrm{s} (10^{-22}\,\mathrm{eV}/m_a)$).} results in different angular frequencies for left and right circularly polarised waves 
\begin{align}
    \omega_{\pm}\simeq k \pm \frac{1}{2}g_{a \gamma}\left(\partial_{t}a+\nabla a \cdot\widehat{\mathbf{k}}\right),
\end{align}
to lowest order in $g_{a\gamma}$.Therefore, the ALP medium is birefringent and  leads to a phase shift between the polarisation components given by
\begin{align}
    \Delta \phi =\frac{g_{a\gamma}}{2} \int_{t_{s}}^{t_{o}} \frac{da}{dt} dt= \frac{g_{a\gamma}}{2}(a_o-a_e)\equiv\frac{g_{a\gamma}}{2}\Delta \phi. \label{eq:phase}
\end{align} 
 In the previous integral $t_o$ refers to the time of observation, while $t_s$ represents the time at emission (source term). This effect is independent of the frequency of the EM wave and depends only on the ALP field values at the source and the observer~\cite{Carroll:1989vb,Harari:1992ea}.

Ignoring the EM backreaction term in equations of motion arising from Eq.~(\ref{eq:Lagrangian}), the ALP field is described by plane waves that at any time $t$ and position $\vec{x}$ is
\begin{align}
    a(\vec{x}, t) = a_0(\vec{x})\cos\left(m_a t + \delta(\vec{x})\right),
\end{align}
where $a_0(\vec{x})$ and $\delta(\vec{x})$ are the amplitude and phase of the field, respectively, at $\vec{x}$. This description holds for time intervals shorter than the coherence time $\tau_c = (m_a \sigma^2)^{-1}\simeq 10^{13}\textrm{s}~(10^{-22}\textrm{eV}/m_a)(200~\textrm{km/s}/\sigma)^2$, where $\sigma$ is the dispersion velocity of dark matter, and for distances smaller than the coherence length $l_c = (m_a \sigma)^{-1}$. Within these limits, the distribution of ULDM behaves as a coherent field oscillating at the nominal Compton frequency
\begin{equation}
\label{eq:ComptonFreq}
\nu  = \frac{m_a}{2\pi}.
\end{equation}
The amplitude of the ALP field is related to the local DM density but its nature is intrinsically stochastic. In a given coherence patch, labelled by $i$, the ALP amplitude is~\cite{Foster:2017hbq,Centers:2019dyn,Castillo:2022zfl},
\begin{equation} \label{eq:alpha}
a_{0,i}=\frac{\sqrt{2\rho_i}}{m_a}\alpha_i,
\end{equation}
where $\rho_i$ is the average DM density in the patch and with $a_{0,i}$ following a Rayleigh distribution. 

We now consider the time-dependent change in the polarisation of light emitted by pulsars, induced by the ALP birefringent medium, as measured on Earth~\cite{Castillo:2022zfl} (cf. Eq.~\eqref{eq:phase}), 
\begin{align}
\Delta \phi(t) &= \frac{g_{a\gamma}}{\sqrt{2}m_a} \Big[ \sqrt{\rho_o} \alpha_o \cos(m_a t + \delta_o)\nonumber\\
&~~~~~~~~~~~~~~- \sqrt{\rho_s} \alpha_s \cos(m_a (t - T) + \delta_s) \Big],
\label{eq:full_sig}
\end{align}
where $t$ is the local time since the beginning of observations ($t=0$), $T$ is the light travel time from a pulsar to Earth, and $\delta_i$, $\rho_i$, and $\alpha_i$ are the phases, DM densities, and stochastic amplitudes at the source ($i=s$) or observer ($i=o$) patches, respectively. 

\begin{figure}
    \centering
\includegraphics[width=\columnwidth]
{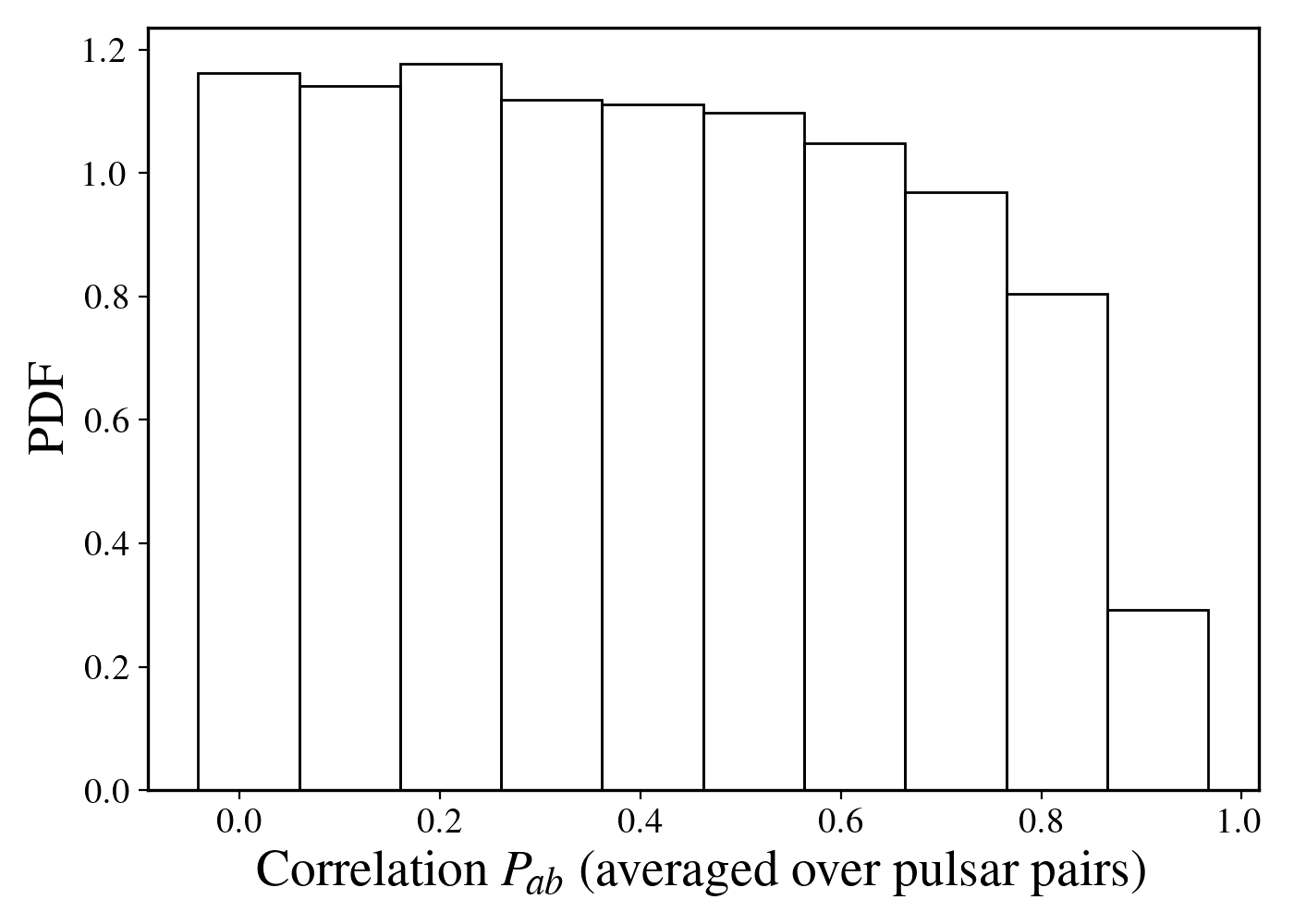}
\caption{Distribution of correlation coefficients averaged over multiple pulsar pairs. Different realizations of the Universe were generated using Eq.~(\ref{eq:full_sig}). The parameters $\alpha_0$ and $\alpha_s$ were sampled from a Rayleigh distribution, while $\delta_0$ and $\delta_s$ were assumed to be uniformly distributed from $0$ to $2\pi$.}
\label{fig:interpsr_corr}
\end{figure}

The distribution of time-averaged inter-pulsar correlation coefficients is given by:
\begin{equation}
P_{ab} = \overline{\Delta \phi(t)_a \Delta \phi(t)_b} = \frac{1}{T_\textrm{tot}}\int_{-T_\textrm{tot}/2}^{T_\textrm{tot}/2}\textrm{d}t \Delta \phi(t)_a \Delta \phi(t)_b\,,
\end{equation}
where $a$ and $b$ are pulsar labels and $T_\textrm{tot}$ is the total timespan of observations. In contrast to the case of gravitational waves (GWs) with inter-pulsar correlation pattern described by the Hellings-Downs curve \cite{1983ApJ...265L..39H}, the ALP signal is expected to exhibit monopolar or fractional monopolar correlation, depending on the configuration of the Earth-pulsar systems. In particular, it is easy to show that if $\alpha_0 \gg \alpha_s$, signals for different pulsars are fully monopole-correlated. In Fig.~\ref{fig:interpsr_corr} we construct the distribution of PTA-averaged inter-pulsar correlation coefficients $P_{ab}$ over multiple universe realisations. As one can see, there is a significant fraction of universes for which the PTA-averaged correlation coefficients are close to 1. However, there are also a number of universes, for which the signal between different pulsars is completely uncorrelated. In order to accelerate the analysis, we perform \textit{incoherent} search and neglect any possible inter-pulsar correlation. In \cite{PhysRevLett.130.121401} authors suggested a method to take into account fractional monopolar correlations. The performance of these two approaches will be compared in future studies.

The right hand side of Eq.~(\ref{eq:full_sig}) can be rearranged in a more compact form:
\begin{equation}
\label{eq:polALP2}
\Delta \phi(t) = \phi_a \cos(m_a t + \varphi_a),
\end{equation}
where
\begin{equation}
\label{eq:phi0}
\phi_a = \frac{g_{a\gamma}}{\sqrt{2}m_a} \left( \rho_o \alpha_o^2 + \rho_s \alpha_s^2 - 2 \sqrt{\rho_o \rho_s} \alpha_o \alpha_s \cos\chi \right)^{1/2},
\end{equation}
and $\chi = m_a T + \delta_o - \delta_s$. Given the large uncertainties on pulsar distances, $m_a T$ spans several cycles of the ALP phase. Hence, hereafter we assume that $\chi$ as well as $\varphi_a$ are distributed uniformly in [$0$, $2\pi$]. In addition, we assume that the average value of the DM density at Earth and at the pulsar are the same
and equal to the measured local DM density, $\rho_0=\rho_s=0.4$ GeV cm$^{-3}$ \cite{2012ApJ...756...89B, 2014JPhG...41f3101R, 2018MNRAS.478.1677S, 2020JPhCS1468a2020D}. Eq.~(\ref{eq:polALP2}) is the final expression that describes the form of the signal to be searched for in the polarisation measurements of pulsars.

\section{Observations and data reduction}
\label{sec:obs_epta}
The EPTA Collaboration has been continuously monitoring $\sim100$ millisecond pulsars (MSPs) for over 25 years \citep{dcl+2016}. The observations are performed using the six largest telescopes in Europe: the Effelsberg 100-m radio telescope (EFF) in Germany, the 76-m Lovell telescope at Jodrell Bank Observatory (JBO) in the United Kingdom, the large radio telescope operated by the Nan\c{c}ay Radio Observatory (NRT) in France, the 64-m Sardinia Radio Telescope (SRT) operated by the Italian National Institute for Astrophysics (INAF) through the Astronomical Observatory of Cagliari (OAC), and the Westerbork Synthesis Radio Telescope (WSRT) operated by ASTRON, the Netherlands Institute for Radio Astronomy. 

The EPTA experiment is primarily designed to detect GWs in the nHz frequency band using pulsars as natural clocks, relying on the technique of high-precision pulsar timing \cite{1978SvA....22...36S, 1979ApJ...234.1100D}. Within this technique, TOAs of pulses from a pulsar are obtained by comparing the observed total intensity profile to a smoothed template profile, using a cross-correlation algorithm. Although pulsar timing traditionally utilises only the Stokes~I or total intensity profiles to obtain the TOAs of the pulses\footnote{The matrix template matching (MTM) technique \cite{straten2006} has been implemented in the \texttt{PSRCHIVE} software to perform TOA extraction using all four Stokes parameters. For instance, MTM has been utilised for TOA extraction for the NUPPI observations.}, all six EPTA instruments have recorded full polarimetry data, such that all four Stokes parameters are available for further analysis. 

The first data release (DR1) of the EPTA \citep{dcl+2016} includes the timing data from 42 MSPs observed between 1996 and 2014. In the summer of 2023, the EPTA collaboration published the second data release (hereafter DR2, \cite{2023A&A...678A..48E}), covering the period from 13.6 to 24.7 years. Due to the large number of telescopes and backend combinations, as well as the diverse data acquisition and processing schemes across the EPTA observatories, 25 MSPs were optimally selected among those monitored by the collaboration, based on the method described in~\cite{2023MNRAS.518.1802S}. Prior to TOA calculation, the raw data are folded modulo the pulsar spin-period with the \texttt{dspsr} software \cite{vanStraten2011}. These folding-mode data cubes undergo a stage of standard initial processing using the \texttt{PSRCHIVE} software \cite{Hotan+2004}, which includes the removal of radio frequency interference (RFI) and polarisation calibration. The latter accounts for the change of parallactic angle during the observation, the geometry of the feed and possible cross-coupling between the two feeds of the receptor (see, e.g. \cite{2001PASP..113.1274H, 2012hpa..book.....L}). Missing or inaccurate calibration will inevitably introduce temporal variations in the polarisation properties of pulsar light, which will cause unwanted artefacts and reduce the overall sensitivity of the dataset. For this reason, an accurate polarimetric calibration is of vital importance for this and similar studies. 

For the current analysis, we make use of calibrated full-Stokes data, which were pre-analysed as part of the EPTA DR2 and subsequently utilised to produce TOA time series. We have limited our analysis to pulsar observations collected by the radio telescopes with the largest effective collective area, namely Effelsberg and Nan\c{c}ay. Our final dataset comprises twelve MSPs with the highest signal-to-noise (S/N) in linear polarisation $L$, with fractional linear polarisation greater than 20. The details of the observational set-up and the steps that have been taken as part of the initial pre-analysis are briefly described in the following paragraphs.

\subsection{Effelsberg data}

The 100-m Effelsberg radio telescope is operated by the Max Planck Institut f\"{u}r Radioastronomie. The observations for the EPTA project are carried out with monthly cadence at three widely distributed frequency bands: 1.4~GHz with P200mm and P217mm receivers, 2.6~GHz with a S110mm receiver and 4.85~GHz with a S60mm receiver (all summarised in \cite{lkg+2016}). The specified integration time per source is 28 min followed by 2 min of noise diode scan. Since 2011 the ROACH-based (Reconfigurable Open Architecture Computing Hardware) backend PSRIX \cite{lkg+2016} digitises the electric signal from the telescope and applies a polyphase filterbank. The \texttt{CoastGuard} pipeline, which is part of the \texttt{Toaster} software library, handles the subsequent necessary steps of the raw data processing \cite{lkg+2016}, i.e.~the collected pulsar observations are coherently dedispersed and folded modulo pulsar period. The resultant data cubes are stored in the \texttt{psrfits} data format with 10-sec and 1024-bin resolution in time and pulse phase, respectively. The RFI-contaminated profiles are identified and erased using \texttt{CoastGuard's} \texttt{surgical} algorithm. The data are further calibrated for flux and polarisation. All of the receivers used for the current study have circularly polarised feeds. Polarisation calibration of cleaned data is performed using the `SingleAxis' technique described in \cite{Britton2000}. Making use of noise diode scans, the relative gain and phase difference of the two polarisation channels have been adjusted. For more details on the initial stages of raw data processing, see \cite{2023A&A...678A..48E}. Polarisation-calibrated data are available from early 2013 to October 2020 and have been used in the current study.

\subsection{Nan\c{c}ay data}
The Nan\c{c}ay Radio Telescope (NRT) is a meridian transit-type instrument with a collective area equivalent to that of a 94-m parabolic dish, making its sensitivity comparable to that of the Effelsberg radio telescope. It is equipped with cryogenically cooled receivers, covering the [1.1, 1.8]~GHz and [1.7, 3.5]~GHz frequency ranges. Regular pulsar observations for the EPTA with the NRT began in 2004. Since August 2011, pulsar observations with the NRT have been conducted with the NUPPI backend (for further details, see \cite{2022A&A...667A..79D, 2013sf2a.conf..327C}), which uses Graphics Processing Units (GPUs) to coherently dedisperse and fold data from the receiver in real time over a total bandwidth of 512~MHz. For the work presented in this article, we considered pulsar observations conducted with the low-frequency receiver at a central frequency of 1484~MHz, which represents the bulk of NRT pulsar observations. The duration of these observations ranges from 20 to 80 mins, and the average observing cadence is $\sim 2$ days. As with the Effelsberg observations, the frequency channels and time sub-integrations heavily contaminated by RFI have been removed. 

Until late 2019, NUPPI pulsar observations were polarisation calibrated with a simple calibration scheme: the calibration was done using observations of a noise diode conducted before each pulsar observation and using the \texttt{SingleAxis} method of \texttt{PSRCHIVE}, which assumes that the polarisation feeds are perfectly orthogonal and that the noise diode signal is 100\% linearly polarised and illuminates the two feeds equally and in phase. Since late 2019, observations of bright, highly linearly polarised pulsars have been conducted regularly, approximately monthly, in a special mode where the horn is made to rotate across the observation. These observations, analyzed with the \texttt{Reception} model of \texttt{PSRCHIVE} enable the determination of the full polarimetric response of the NRT. For the work presented here, we used NRT pulsar archives recorded at L-band from October 2019 to April 2023, i.e., calibrated with this much-improved polarisation calibration scheme. Further details about the NRT, pulsar observations with the NRT and the polarisation calibration of the NRT data can be found in \cite{guillemot2023}.

%

\subsection{Ionospheric correction}\label{sec:IonosCorr}
\label{sec:iono_correction}

The effect of ULDM is not the only process that can alter the polarisation properties of pulsar emission. In fact, the dominant contribution comes from the Faraday rotation that takes place in the magneto-ionic plasma between the source and the observer, causing the rotation of the plane of polarisation of the linearly polarised part of the radiation. In astronomy, the orientation of the polarisation plane is characterised by the position angle (PA):
\begin{equation}
\textrm{PA} = \Psi_0 + \text{RM}\, \lambda^2,
\label{eq:farrot}
\end{equation}
where angle $\Psi_0$ characterises the initial orientation of the plane of polarisation (as emitted by a source) and $\lambda$ is the observing wavelength. The rate at which the plane of polarisation rotates is characterised by the \textit{rotation measure} (RM):
\begin{equation}
\textrm{RM} = 0.81 \mathrm{\left[\frac{rad}{m^2}\right]} \int_{\text{LoS}}n_e(\textbf{B} \cdot \text{d}\textbf{l}),
\label{eq:rmint}
\end{equation}
where $n_e$ (cm$^{-3}$) is the electron number density of the intervening plasma and the dot product in brackets represents the projection of the magnetic field $\textbf{B}$ (measured in $\mu$G) onto the infinitesimal interval d$\textbf{l}$ of the distance along the line of sight (LoS) expressed in pc. 

As can be seen from Eq.~(\ref{eq:farrot}), Faraday rotation is strongly dependent on the observational frequency, whereas the effect of ultra-light axions is independent of the EM wave frequency (the signal is achromatic), allowing the two to be disentangled. However, that would only be possible for the observing systems with large fractional bandwidth\footnote{Fractional bandwidth is the ratio of the total observed bandwidth to the central frequency of operation.}, so that the plane of polarisation rotates considerably along the bandwidth. For instance, for the Nan\c{c}ay L-band receiver the relative change in PA between the upper and lower part of the band is only $\sim2$ deg, which can not be probed with the current sensitivities. Therefore the two signals due to the ULDM and Faraday rotation in the intervening plasma, have similar manifestation in the data for the given observational setup. 

As we are not sensitive to time-independent modifications of the PA (given that the initial polarisation configuration is unknown), only time-variable effects should be considered in the analysis. Generally, there are two known chromatic, i.e. which depend on observational radio frequency, processes that can alter the polarisation properties of pulsar radiation dynamically. The main contributor to variations in RM and hence PA is the terrestrial plasma, i.e. ionosphere and plasmasphere \cite{2012RaSc...47.0K08O, 2013A&A...552A..58S, pnt+2019}. The second type of temporal RM fluctuations is driven by spatial turbulence in the ionised interstellar medium (IISM), over which the LoS scans as a pulsar moves in the tangential plane. However, the latter is estimated to be five orders of magnitude lower than the former and cannot be measured with current sensitivities (see more details in \cite{pnt+2019}). Therefore, only the atmospheric contribution is accounted for in the current analysis. 

The ionosphere is modelled using the conventional single-layer model (SLM), which assumes that the terrestrial plasma is encapsulated in an infinitesimally thin slab fixed at a height of $\sim450$~km. Within this approximation, Eq.~(\ref{eq:rmint}) reduces to a simple product of the LoS projected geomagnetic field $\textrm{B}_\textrm{E}$ and integrated electron density, also known as \textrm{slant total electron content} (STEC). Both values are evaluated at the ionospheric pierce point (IPP), where the LoS intersects the ionospheric layer. Information about the Earth's magnetic field is provided by the World Magnetic Model \cite{wmm2020}. The STEC values are recalculated from the vertical TECs (VTECs), which are electron column densities in the direction of the zenith. VTECs are extracted from the interpolated global ionospheric maps (GIMs), which contain VTEC values defined on a regular temporal and spatial grid reconstructed from the Global Positioning System (GPS) observations. Specifically, for the current work, we have used UQRG GIMs produced by the UPC-IonSat collaboration \cite{1998RaSc...33..565M, 2009JGeod..83..263H}. The UQRG GIMs were shown to surpass other publicly available GIMs when comparing different ionospheric models using the observed Faraday rotation of pulsars at low frequencies of LOFAR \cite{2019MNRAS.483.4100P}. However, the authors also showed that none of the existing models, including UQRG, have been able to capture the full complexity of ionospheric physics. In particular, it was found that two types of systematics emerge: quasi-sinusoidal variations with a one-year period, and long-term linear trends strongly correlated with the 11-yr Solar cycle. See more details in \cite{2019MNRAS.483.4100P, 2023JGeod..97..116P}. The possible presence of this interference was taken into consideration within a Bayesian framework (see Section~\ref{sec:bayes_analysis}).

\begin{figure}
    \centering
\includegraphics[width=\columnwidth]
{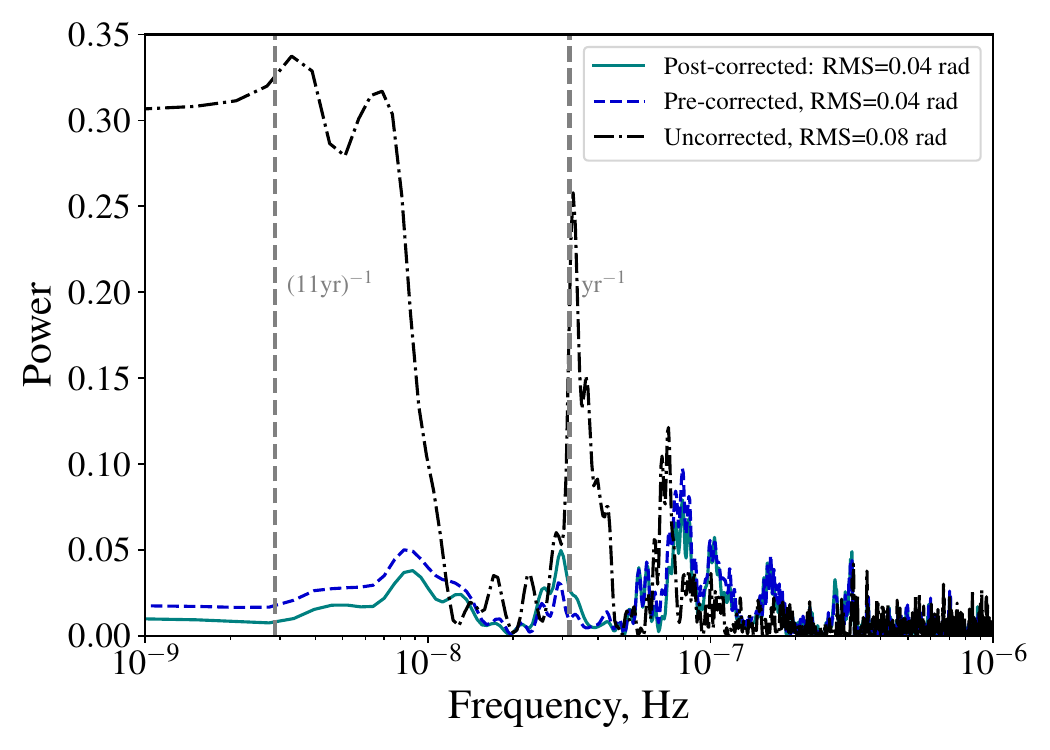}
\caption{Lomb-Scargle periodogram of the PA residuals for PSR~J1600$-$3053. The spectra of $\Delta$PA, when the ionospheric contribution has been subtracted, before (pre-processing scheme) and after (post-processing scheme) PA extraction are shown in blue (dashed) and turquoise (solid), respectively. The Lomb-Scargle periodogram of $\Delta$PA uncorrected for the ionospheric effect is shown in black (dot-dashed).}
\label{fig:iono_corr}
\end{figure}

The ionospheric correction was implemented at each fully frequency-resolved pulsar archive (before the stage of PA extraction) by modifying the relevant entry in the \texttt{psrfits} header using \texttt{--aux\_rm} command. In this case, the possibly present (weak) effect of depolarisation due to the ionosphere, which can alter the shape of the polarisation profile and bias the PA measurements, is compensated. It is worth noting that we have also tested the performance of the ionospheric correction in the post-processing stage (after PA extraction), neglecting the frequency dependence of the Faraday rotation within an observing band. As expected, for instruments with small fractional bandwidths that we are using in the current project, the difference in the resultant standard deviation of the residuals for the two schemes is negligible (see Fig.~\ref{fig:iono_corr}).

\subsection{PA extraction}

The astrophysical signal of interest manifests itself as a periodic variation of the PA (see Eq.~(\ref{eq:full_sig})). The following steps are undertaken to capture these changes in PA relative to a baseline. We start by correcting the pulsar data, stored in \texttt{psrfits} format, for the ionospheric contribution, as described in Sec~\ref{sec:iono_correction}. Afterwards, we build a template by summing up a number of observations with a high S/N ratio, followed by the smoothing. We force-align averaged pulse profiles taken at different observational epochs in phase, using the Stokes~I template constructed on a previous step as a benchmark. This procedure is necessary as multiple stochastic processes, primarily the intrinsic spin noise, can phase-shift the pulse profiles and bias the final measurements of the PA (as PAs of an individual archive and of a template generally may not match the same phase). Finally, we calculate the difference between the PA across an individual profile and a template.
The phase-resolved PA residuals are constructed directly from Stokes~$Q$ and $U$ observables, using the following expression:
\begin{equation}
\begin{split}
\Delta \textrm{PA}(\phi) = \frac{1}{2}\left(\arctan\frac{U_{\textrm{epoch}}}{Q_{\textrm{epoch}}} - \arctan\frac{U_\textrm{tmpl}}{Q_\textrm{tmpl}}\right)=\\
\frac{1}{2}\arcsin\{U_{\textrm{epoch}}(\phi)Q_{\textrm{tmpl}}(\phi)-Q_{\textrm{epoch}}(\phi)U_{\textrm{tmpl}}(\phi)\},
\label{eq:delta_pa_qu}
\end{split}
\end{equation}
where $Q_{\textrm{epoch}}/U_{\textrm{epoch}}$ and $Q_{\textrm{epoch}}/U_{\textrm{tmpl}}$ are Stokes~Q and U profiles as a function of a pulse phase $\phi$ of an individual observation and the template, respectively. The sought $\Delta \textrm{PA}$ is a weighted average of $\Delta$PA($\phi$), where we use the inverse square of measurement uncertainties $1/\sigma_{\textrm{PA}}^2$ as weights:
\begin{equation}
\sigma_{\textrm{PA}}=\frac{\sqrt{Q_{\textrm{epoch}}^2(\phi)\sigma_U^2 + U_{\textrm{epoch}}^2(\phi)\sigma_Q^2}}{2(Q^2_{\textrm{epoch}}(\phi) + U^2_{\textrm{epoch}}(\phi))}.
\end{equation}
Here, $\sigma_Q$ and $\sigma_U$ are the off-pulse root-mean-squares. The average uncertainty on the $\Delta$PA is calculated as the standard deviation of $\Delta$PA$(\phi)$ normalised by the number of on-pulse measurements.
\begin{figure*}
    \centering
\includegraphics[width=\textwidth]
{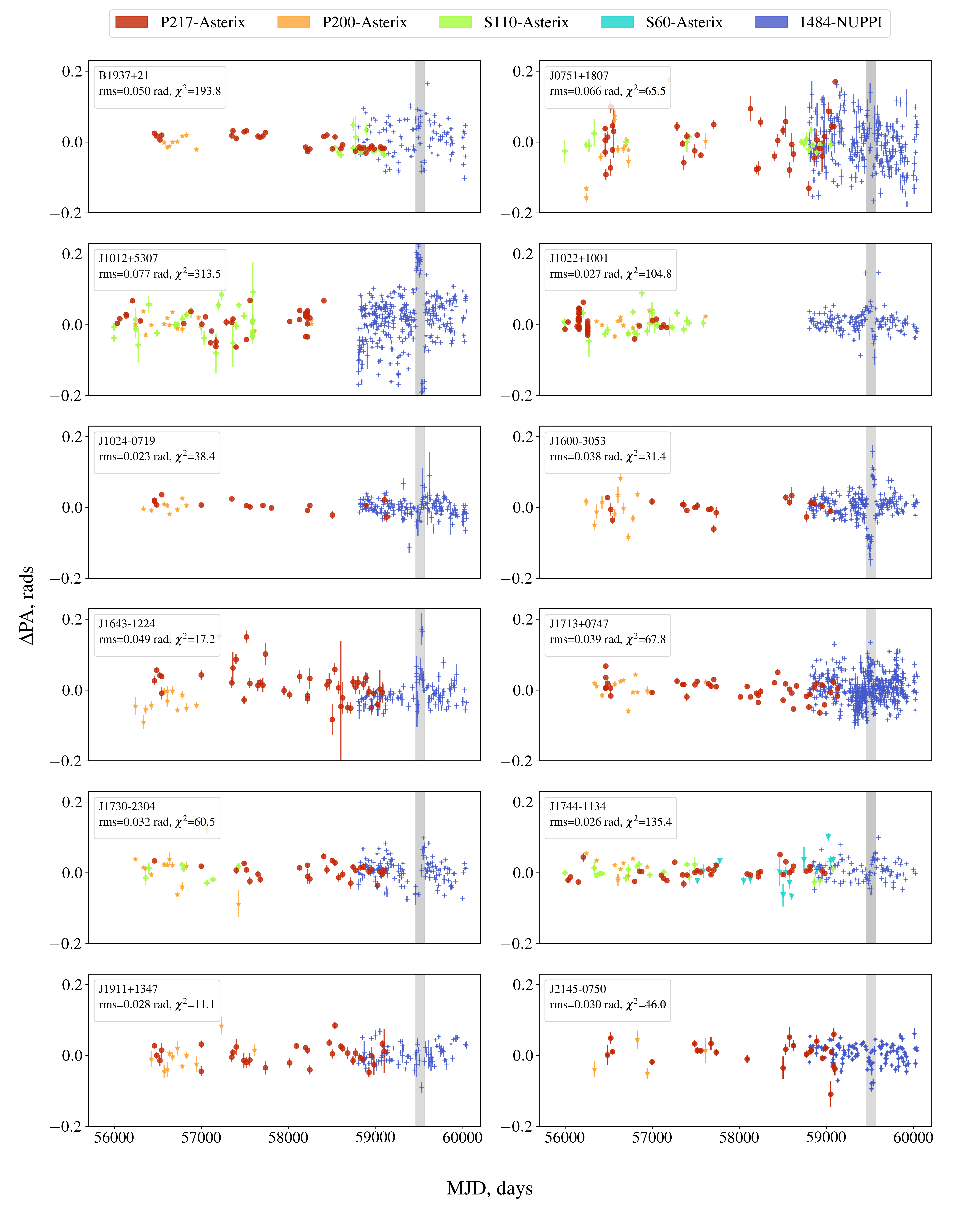}
\caption{The PA residuals of the subset of twelve pre-selected EPTA DR2 MSPs. Multiple colours correspond to four different observing systems: P217-Asterix (red circles), P200-Asterix (orange stares), S110-Asterix (green filled pluses), S60-Asterix (turqouise triangles), 1484-NUPPI (blue pluses). The shadow area, which extends from MJD 59460 to 59560, represents the malfunction of one of the local oscillators at the Nan\c{c}ay observatory. One of the consequences of this problem are the jumps in PA residuals, which are particularly apparent in the data of PSR~J1012+5307, J1643$-$1224 and J1600$-$3053. The $\chi^2$ values indicated at the panels are calculated after eliminating the epochs affected by the malfunction issue.}
\label{fig:pa_total}
\end{figure*}

\begin{figure*}
    \centering
\includegraphics[width=\columnwidth]
{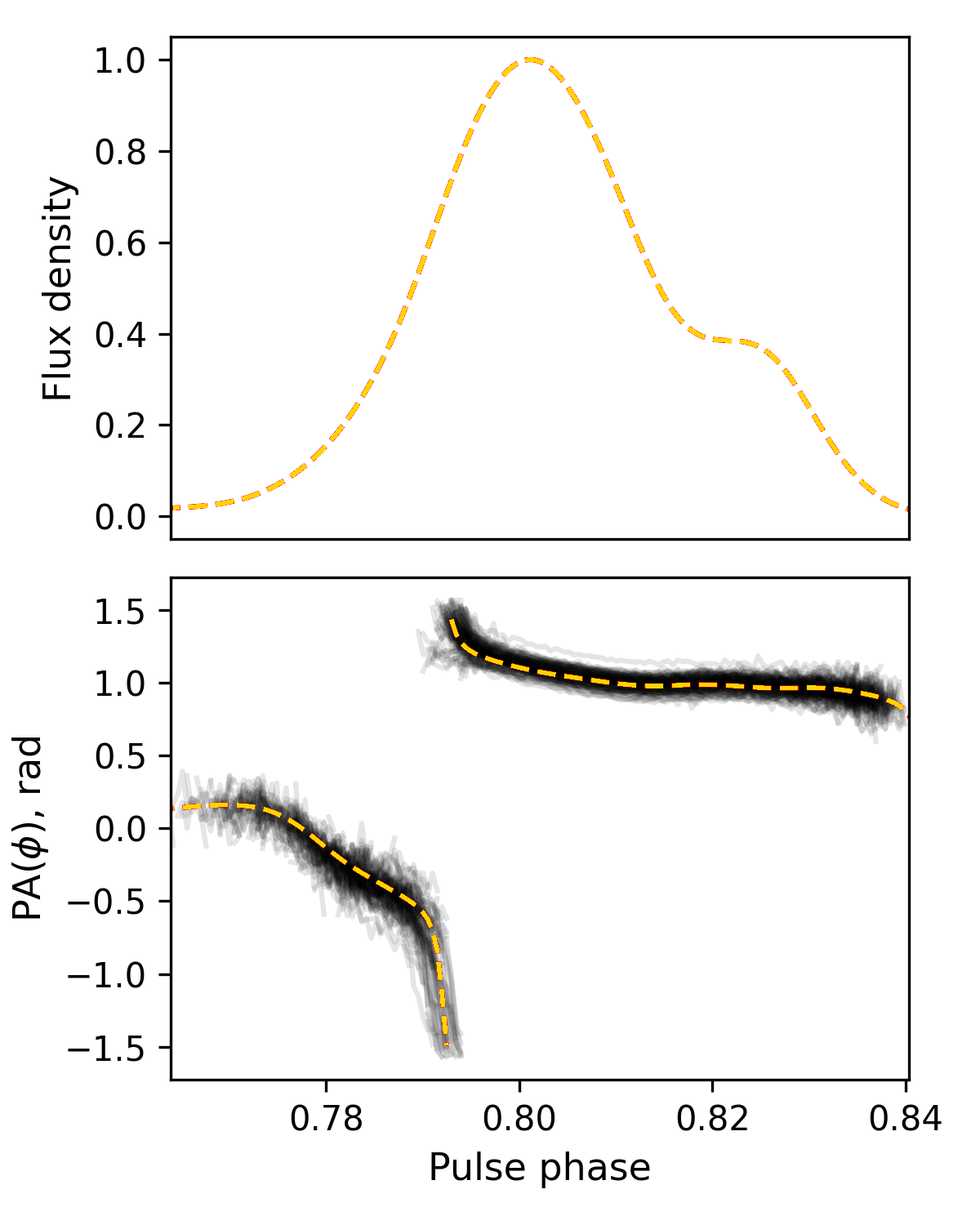}
\includegraphics[width=\columnwidth]
{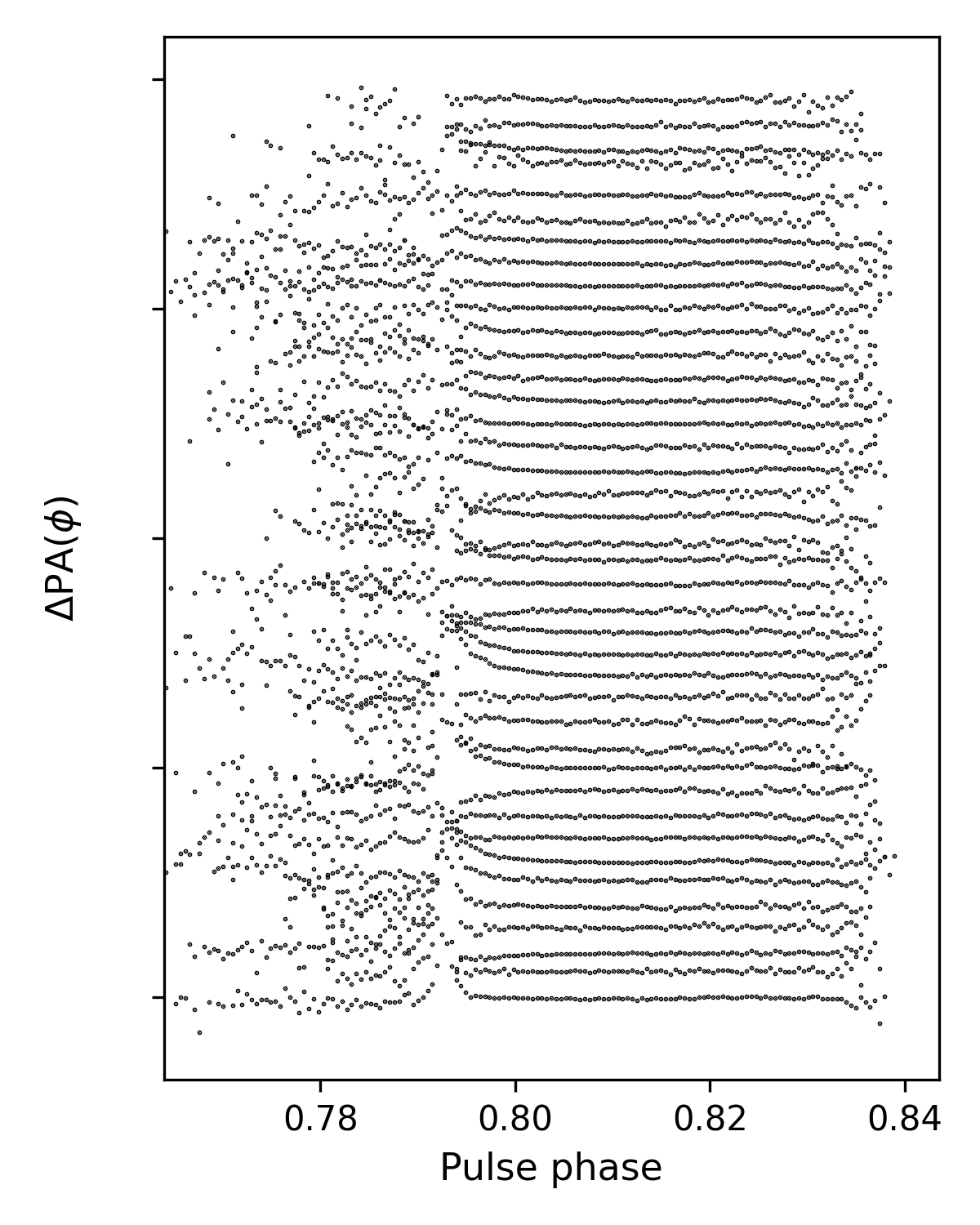}
\caption{Intensity and polarisation configurations of PSR~B1937$+$21. On the left: template intensity profile (upper panel) and individual PA$(\phi)$ profiles in black overlapped with the template PA profile in yellow (lower panel). On the right: waterfall plots of the $\Delta$PA$(\phi)$ observed with NRT. Each line corresponds to a specific epoch of observation. The red line on the left panel is the PA swing of the smoothed template. For the sake of presentation, only forty epochs have been plotted. It is clear that $\Delta$PA$(\phi)$ exhibits non-Gaussian systematic behaviour. These observed features are particularly pronounced in the regions close to the orthogonal jump of the PA.}
\label{fig:residual_waterfall}
\end{figure*}

The obtained dataset is publicly available on zenodo\footnote{\protect\url{https://doi.org/10.5281/zenodo.10.5281/zenodo.1400595}} and is displayed in Fig.~\ref{fig:pa_total}. The reported $\chi^2$ values indicate that the uncertainties are considerably underestimated. A similar behaviour has been reported previously by the Parkes PTA Collaboration \citep[PPTA][]{ymv+2011}. This bias is caused by the presence of systematics (i.e., deviations from non-Gaussianity) in phase-resolved PA residuals, calculated using Eq.~(\ref{eq:delta_pa_qu}). An example of such residual systematics between the epoch-wise and template PA measurements for PSR~B1937$+$21 is shown with the waterfall plot in Fig.~{\ref{fig:residual_waterfall}}. We expect that such systematics may be caused by the scattering in the IISM, or errors in polarisation calibration, or other instrument-related factors, and will be further explored in future work (see Sec.~\ref{sec:conclusion} for discussion).

\section{Methods}
\label{sec:method}
In this section, we introduce the methods used to perform parameter estimation and search for the ULDM signature in the data using frequentist and Bayesian inferences. For both types of analyses, we assume that the Earth and pulsar terms fall in different coherent patches and that the monochromatic, i.e.~sinusoidal with period unambiguously defined by the mass of the ALP, signal from ULDM is uncorrelated between different pulsars. For our sample of pulsars, this assumption holds for $\geq 5\times 10^{-23}$~eV. In App.~\ref{sec:app_stochasticity} we further discuss the justification for this assumption and investigate how upper limits are modified for different pulsar-Earth configurations.
We also treat the noise in the PA residuals as white Gaussian. In the Bayesian framework, we additionally take measures to account for the underestimation in the uncertainties and presence of long-term systematics with white noise modifiers, an extra error in quadrature (EQUAD), which represents an additional unaccounted noise, and an error scaling factor (EFAC) \cite{2014MNRAS.437.3004L}, and inclusion of linear and quadratic trends in the model, respectively.
\subsection{Frequentist analysis}\label{sec:Freq}

The frequentist approach involves searching for periodic signals in unevenly distributed time series with the Lomb-Scargle (LS) periodogram~\cite{Lomb:1976wy,Scargle:1982bw}, based on a combination of signal analysis techniques~\cite{VanderPlas:2018rru}. 
In this work we employ the \textit{generalised} LS periodogram (GLSP)~\cite{Zechmeister:2009js,VanderPlas:2018rru} $P_\textrm{LS}(\nu)$ and follow the procedure described in~\cite{Castillo:2022zfl}.

First, we look for peaks in $P_\textrm{LS}(\nu)$ and quantify their significance by comparing them with the so-called false alarm probability (FAP). This value quantifies the probability that a significant peak emerges from purely random fluctuations, and we estimate it using the \textit{bootstrap} method, which resamples the time series while keeping the temporal coordinates fixed using Monte Carlo simulations~\cite{VanderPlas:2018rru}. The highest peaks of these simulations form a probability distribution that is used directly to compute the FAP.

After verifying that there are no peaks compatible with the expected DM signal, we perform $10^4$ Monte Carlo simulations, injecting a harmonic signal into the time series,
\begin{equation}
    \Delta \phi_{\rm{sim}} = \phi \cos\left(2\pi\nu\, t + \varphi\right),
    \label{eq:harm_osc}
\end{equation}
with a given frequency $\nu$, amplitude $\phi$, and a random phase $\varphi$ uniformly sampled from $[0,2\pi]$. Using the GLSP, we compute the probability density function (PDF) of $P_\textrm{LS}(\nu,\phi)$ and determine $\phi_{95}$ such that $P_\textrm{LS}(\nu,\phi_{95})$ matches the experimental $P_\textrm{LS}(\nu)$. This provides an upper bound on $\phi$ at 95\% confidence level (C.L.). We use a rolling mean method to smooth out the fluctuations, averaging $\phi_{95}$ within a window centred at each frequency. This process is repeated for all frequencies and pulsars.

Finally, we perform a combined analysis, conducting a Monte Carlo computation similar to one in the previous section. Using the PDF of $\phi_a$ obtained earlier, we generate $10^5$ pseudo-experiments, drawing values of $\phi_{a,i}$ for each pulsar, along with values for $\alpha_o$, $\alpha_{s,i}$, and $\chi_i$ for the pulsars. Using Eq.~(\ref{eq:harm_osc}), we calculate $g_{a\gamma,i}$ for each pulsar. The global value of $g_{a\gamma}$ is obtained as a weighted mean, with weights determined by propagating the root squares of the variances of $\phi_{a,i}$, directly related to the scale factor of the Rayleigh distribution. By running multiple pseudo-experiments, we construct its PDF and report the upper limit on the axion-photon coupling at the 95\% confidence level from the resulting global $g_{a\gamma}$ distribution.

\subsection{Bayesian analysis}
\label{sec:bayes_analysis}
In the Bayesian framework we assume that the $\Delta$PA measured at observational epoch $t_i$, $\Delta\textrm{PA}^{\text{obs}}(t_i)$, for a single pulsar is a sum of an astrophysical signal of interest $\delta\phi^\text{ULDM}(t_{i})$, given by Eq.~(\ref{eq:polALP2}), and a Gaussian white noise $n (t_{i})$. Given the flexibility of the Bayesian scheme in comparison to the GLSP, we complement the frequentist approach by considering the possible presence of multiple deterministic processes $\textrm{PA}^{\text{det}}(t_i)$ which may emerge due to, e.g. polarisation calibration errors or poor ionospheric modelling:
\begin{equation}
\Delta\textrm{PA}^{\text{obs}}(t_i) = \delta\phi^\text{ULDM}(\pmb{\Theta}, t_{i}) + \textrm{PA}^{\textrm{det}}(\pmb{\Psi}, t_i) + n (t_{i}),
\label{eq:pa_obs}
\end{equation}
where $\pmb{\Theta}$ is a parameter vector of the ALP model and $\pmb{\Psi}$ parameterise the deterministic contribution. We assume that $\textbf{PA}^{\textrm{det}}$ linearly depends on $\pmb{\Psi}$, so that it can be expressed as:
\begin{equation}
    \textbf{PA}^{\textrm{det}} = \textbf{M} \pmb{\Psi},
\label{eq:design_M}
\end{equation}
where $\textbf{M}$ is the $(N \times m)$ design matrix of partial derivatives of each parameter characterising deterministic trends ($N$ is the total number of measurements  and $m$ is the dimensionality of $\pmb{\Psi}$). In our work, we consider linear and quadratic trends, as well as sinusoidal variations with 1-yr and 11-yr periods responsible for commonly known biases in the ionospheric modelling \citep{pnt+2019, pmh+2023}. The inclusion of the deterministic term in the model removes possible residual trends unrelated to the ALPs signal, but also provides more conservative estimates of the target parameter $g_{a\gamma}$. Within the Bayesian framework, all systematics, noises and signals of interest are considered simultaneously using the multivariate Gaussian likelihood:

\begin{equation}
\begin{aligned}
    &\mathcal{L} (\pmb{\Delta} \textbf{\textrm{PA}} | \pmb{\Theta}, \pmb{\Psi}) = \frac{1}{\sqrt{|\text{det}2\pi \textbf{C}|}}  \\
    &\times \text{exp} \; \{-\frac{1}{2}(\pmb{\Delta}\textbf{PA}^{\text{obs}} - \pmb{\delta\phi}^\text{ULDM} - \textbf{PA}^{\textrm{det}})^T
    \textbf{C}^{-1}
    \\
    &~~~~~~~~~~~~~~~~~~(\pmb{\Delta}\textbf{PA}^{\text{obs}} - \pmb{\delta\phi}^\text{ULDM} - \textbf{PA}^{\textrm{det}})
    \},
\end{aligned}
\label{eq:gauss_L}
\end{equation}
where $\textbf{C}=\sigma_\textrm{PA}^2\delta_{ij}$ is the covariance matrix of the stochastic processes and $\delta_{ij}$ is the Kronecker delta. Given the misestimation of the uncertainties, two additional parmeters, EFAC and EQUAD, were addede to the analysis, so that the diagonal elements of the covariance matrix are modified as follows: $\textbf{C}_{ii}=\textrm{EFAC}^2\sigma_\textrm{PA, ii}^2+\textrm{EQUAD}^2$. This likelihood function can be marginalised analytically over the parameters $\pmb{\Psi}$ and further reduced to a more compact form \citep{vlm+2009, vl2013}:

\begin{equation}
    \mathcal{L} (\pmb{\Delta} \textbf{\textrm{PA}} | \pmb{\Theta}) = \frac{\text{exp} (-\frac{1}{2}\pmb{\delta}\textbf{PA}^T \textbf{G} (\textbf{G}^T\textbf{C}\textbf{G})^{-1}\textbf{G}^T \pmb{\delta}\textbf{PA})}{\sqrt{(2\pi)^{N-m}\text{det}(\textbf{G}^T\textbf{C}\textbf{G})}},
    \label{eq:G_matrix_L}
\end{equation}
where $\pmb{\delta}\textbf{PA}=\pmb{\Delta}\textbf{PA}-\pmb{\delta\phi}^\textrm{ULDM}$ and the $(N \times N-m)$ matrix $\textbf{G}$ was obtained from the singular-value decomposition of the design matrix $\textbf{M}$. Specifically, $\textbf{M} = \textbf{UDV}^T$, where $\textbf{U}$ and $\textbf{V}$ are $(N \times N)$ and $(m \times m)$ orthogonal matrices, respectively and $\textbf{D}$ is an $(N \times m)$ diagonal matrix with singular values of $\textbf{M}$. The $\textbf{G}$ matrix is formed from the $\textbf{U}$ matrix such that $\textbf{U} = (\textbf{G}_\textbf{c} \; \textbf{G})$. See more details in, e.g. \cite{taylor2021}.

After constructing the likelihood function, the Bayesian analysis proceeds with the determination of the posterior distributions of the parameters $p^{\text{post}}(\pmb{\Theta} | \pmb{\Delta}\textbf{\textrm{PA}})$ by updating the prior distribution of those parameters $p^{\text{prior}}(\pmb{\Theta})$ through the likelihood function $\mathcal{L}(\pmb{\Delta}\textbf{\textrm{PA}}|\pmb{\Theta})$:

\begin{equation}
 p^{\text{post}}(\pmb{\Theta} | \pmb{\Delta} \textbf{\textrm{PA}}) = \frac{\mathcal{L}(\pmb{\Delta} \textbf{\textrm{PA}}|\pmb{\Theta})p^{\text{prior}}(\pmb{\Theta})}{Z},
 \label{eq:bayes}
\end{equation}
where the denominator $Z$ is called \textit{evidence} and can be expressed as:

\begin{equation}
    Z = \int \mathcal{L}(\pmb{\Delta}\textbf{\textrm{PA}}|\pmb{\Theta})p^{\text{prior}}(\pmb{\Theta}) d^n (\pmb{\Theta}).
    \label{eq:evidence}
\end{equation}
Evidence is a key criterion for the hypothesis testing (or model
selection) problem. One can use it to calculate the Bayes factor (BF) to choose between two hypotheses, which in our case are: {\it i)} signal present in the data; and {\it ii)} signal missing in the data. For the task of hypothesis testing, we do not evaluate the above integrals, but use the Savage-Dickey ratio instead, which is defined as the ratio of the posterior to the prior density at the critical value $\pmb{\Theta_0}$, corresponding to the zero amplitude $g_{a\gamma} = 0$:

\begin{equation}
    B_{21} = \frac{p^{\text{post}}(\pmb{\Theta_0} | \pmb{\Delta}\textbf{\textrm{PA}})}{p^{\text{prior}}(\pmb{\Theta_0})}.
    \label{eq:SD}
\end{equation}

\begin{figure}
    \centering
\includegraphics[width=\columnwidth]
{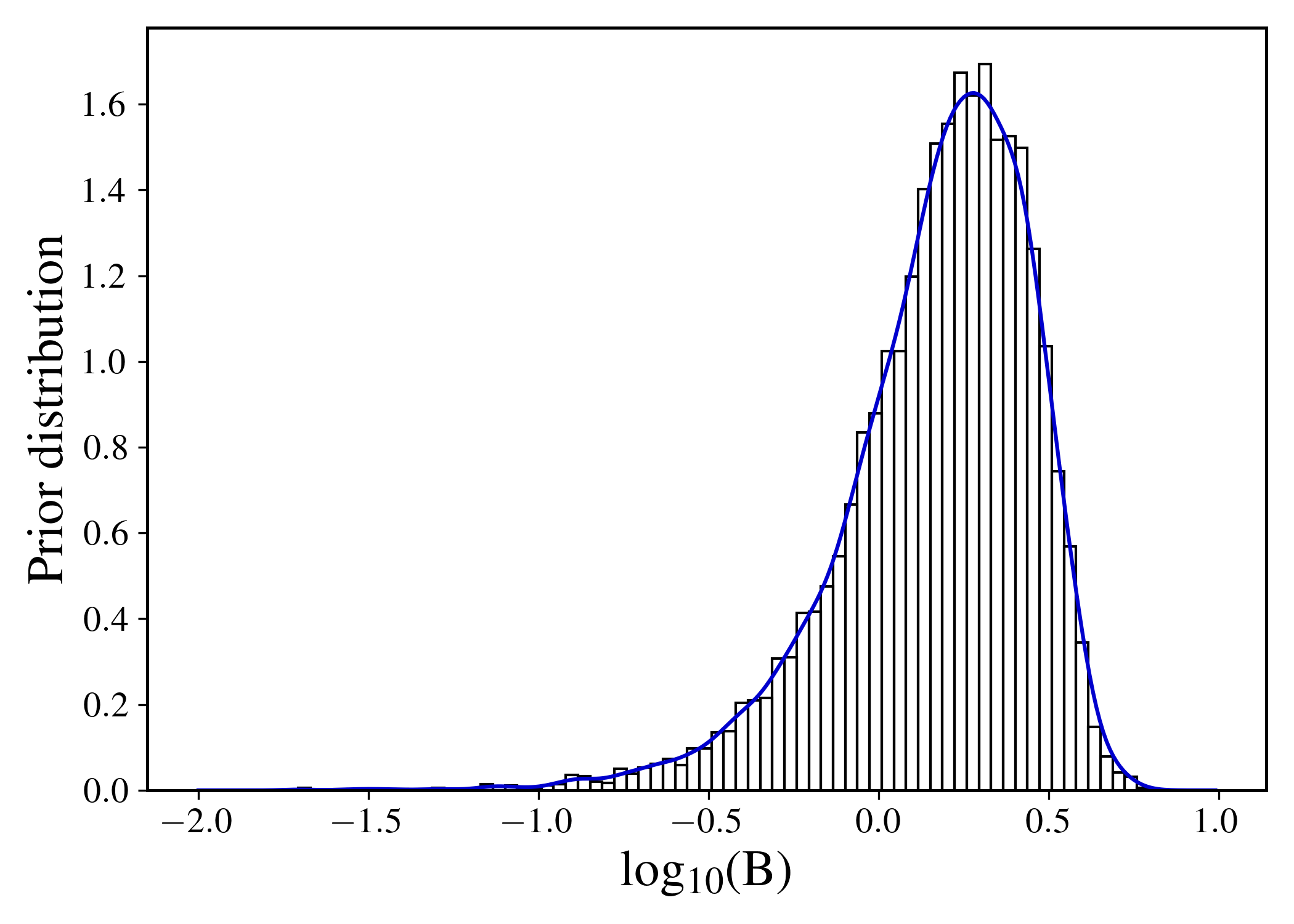}
\caption{Numerically reconstructed prior distribution of the normalisation parameter $B$ and its KDE approximation (in blue).}
\label{fig:bracket}
\end{figure}

Conventionally, the Bayes factors $B_{21}=1$ and $B_{21}=2$ are considered substantial and  evidence, respectively. For the parameter estimation problem, evidence $Z$ plays the role of normalisation constant. In the case of flat uninformative priors, the Eq.~(\ref{eq:bayes}) is reduced to   $p^{\text{post}}(\pmb{\Theta} | \pmb{\Delta} \textbf{\textrm{PA}}) \sim \mathcal{L}(\pmb{\Delta} \textbf{\textrm{PA}}|\pmb{\Theta})$, which is an expression used to construct the posterior distribution of the signal parameters in practice. The model parameters $\pmb{\Theta}$ are split into two categories:
{\it i)} common to all the pulsars, which are the coupling constant $g_{a\gamma}$, and the axion mass $m_a$;
{\it ii)} individual to a pulsar, which are the phase of oscillation $\phi_a$, the normalisation parameter $B=\sqrt{\alpha_o^2 + \alpha_s^2 - 2\alpha_o\alpha_s \cos \chi}$, the error in quadrature EQUADs (one per observing system), the error multiplier EFACs (one per observing system) and the OFFSETs (one per observing system), which are introduced to describe the jumps between different observing systems. The distribution of the parameter $B$ has been estimated numerically using the priors of the relevant input parameters ($\alpha_0$, $\alpha_s$ and $\chi$) and approximated using a kernel density estimator (KDE) (see Fig.~\ref{fig:bracket}). The whole set of parameters and the corresponding priors are summarised in Table \ref{tab:bayes_params}.

To reduce the complexity of the current analysis, we neglect any possible cross-correlation between datasets of different pulsars (see Fig.~\ref{fig:interpsr_corr}). In this case the likelihood in Eq.~(\ref{eq:G_matrix_L}) can be factorised as:
\begin{equation}
\begin{aligned}
\mathcal{L} (\mathbf{\Delta} \textbf{\textrm{PA}} | g_{a\gamma}, m_{a}, \textbf{A}_1, \textbf{A}_2, ..., \textbf{A}_N)
\\
=\prod_{k=1}^{N}\mathcal{L} (\mathbf{\Delta} \textbf{\textrm{PA}} | g_{a\gamma}, m_{a}, \textbf{A}_k),
\end{aligned}
\label{eq:lik_fact}
\end{equation}
where $\textbf{A}_k$ are groups of parameters individual to a pulsar. Using this expression, one can perform an efficient search for an astrophysical signal in a reduced parameter space for each pulsar dataset separately and construct the joint posterior probability in the after-processing. Eq.~(\ref{eq:lik_fact}) is the final expression that we used to perform the Bayesian run.

\begin{table*}
\caption{The set of parameters, used for the Bayesian analysis}
\begin{ruledtabular}
\begin{tabular}{cccc}
 \textbf{Parameter} & \textbf{Description} & \textbf{Prior} & \textbf{Comments} \\
 \hline
  {\it Noise parameters}&&&\vspace{-0.3cm}\\
 \multicolumn{4}{c}{}\\
 EFAC & White-noise modifier per backend & Uniform [0, 4] & One parameter per backend \\
 EQUAD & Additive white noise per backend & log$_{10}$-Uniform [-8, 3] &  One parameter per backend\\
 OFFSET & An offset between observing systems & Uniform [-2, 2] & One parameter per backend \\
 \hline
 {\it Signal parameters}&&& \vspace{-0.3cm}\\
 \multicolumn{4}{c}{} \\
 $g_{a\gamma}$ & photon-axion coupling constant & log$_{10}$-Uniform [-6, 5] (search) & One parameter per PTA \\
 & &Linear-Exp [-7, 6] (upper limits) & \\
 $f$(Hz) & Oscillation frequency & log$_{10}$-Uniform [1/T, $10^{-6}$] & Fixed, regular grid in log-scale \\
 $\phi_a$ & Oscillation phase & Uniform[$0, 2\pi$] &  \\
 $B$ & $B = \sqrt{\alpha_o^2 + \alpha_s^2 - 2\alpha_o\alpha_s \cos \chi}$ & Estimated numerically using KDE & \\
 && (see Fig.~\ref{fig:bracket})&\\
 {\it Additional parameters}&&& \vspace{-0.3cm}\\
 \\
 $\alpha_0$ & Stochastic amplitude on Earth & Rayleigh distribution & $p(\alpha)=\alpha \textrm{exp}\left(-\frac{\alpha^2}{2}\right)$ \\
 $\alpha_s$ & Stochastic amplitude on a pulsar & Rayleigh distribution & $p(\alpha)=\alpha \textrm{exp}\left(-\frac{\alpha^2}{2}\right)$ \\
 $\chi$ & Phase & Uniform[$0, 2\pi$] & $\chi = m_a T + \delta_o - \delta_s$ \\
 \multicolumn{4}{c}{} \\
\end{tabular} 
\end{ruledtabular}
\label{tab:bayes_params}
\end{table*}

\section{Results}
\label{sec:result}
We now turn to the description of the main results of the analysis. We present the results of the frequentist and the Bayesian analysis separately, discussing first the detection and then upper limit estimations.

\subsection{Frequentist analysis}

\begin{figure}
    \centering
\includegraphics[width=\columnwidth]
{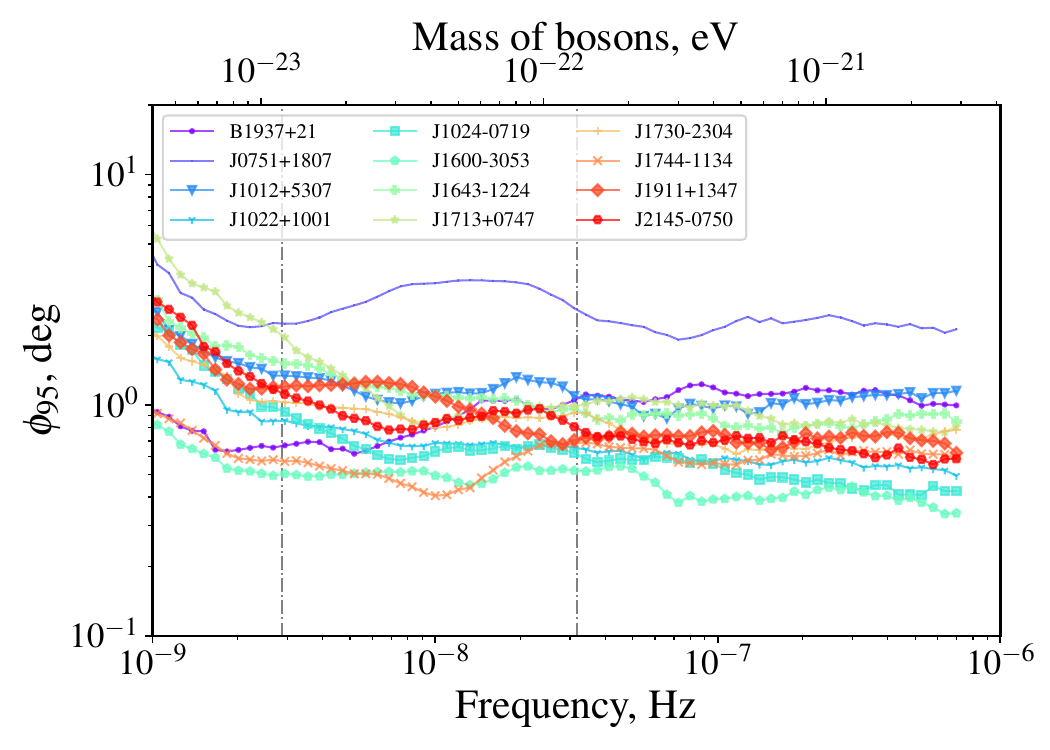}
\caption{\label{fig:freq_upper_lim_phi95} Frequentist limits on the amplitude at the $95\%$ C.L. ($\phi_{95}$)  for the 12 pre-selected EPTA pulsars. These constraints will be used to produce a combined bound directly on $g_{a\gamma}$, which should also include the effects of stochastic amplitudes (so, is not illustrated in the current Fig.).}
\end{figure}

Following the procedure described in Sec.~\ref{sec:Freq}, we first search for peaks in the GLSP. We only find a peak below the 1\% FAP, in PSR~J1600$-$3053, at a frequency of $\sim\ 3\times 10^{-8}\,\rm Hz$ corresponding to a periodicity of a year. The lack of similar peaks in the GLSP of all the other pulsars points to a different explanation for the ULDM case. In fact, the 1-year frequency aligns well with a seasonal ionospheric signal, still remaining after the corrections, see Sec. \ref{sec:IonosCorr}. After confirming that there is no signal compatible with the ULDM, we proceed to set constraints on periodic signals for each pulsar, finding the maximum 95\% C.L. amplitude, $\phi_{95}$. We show them in Fig.~\ref{fig:freq_upper_lim_phi95}.

We see that all pulsars set an upper limit of around a degree in amplitude, approximately constant over a wide range of frequencies, except for the lower frequencies, where the sensitivity drops. In order to translate these results to ULDM parameters, i.e. to the coupling $g_{a\gamma}$ in Eq. \eqref{eq:Lagrangian}, we perform the global analysis. We first asses the local DM density of each source and, to take into account stochastic effects, the distance between all of them to check if there are sources in the same coherence patch. We also produce the results under the assumptions that the local DM density at each source and the Earth is the same, $\rho_{\rm DM}\approx 0.4\ \rm GeV/cm^3$ and neglecting stochastic effects, i.e. setting the stochastic parameter in Eq.~\eqref{eq:alpha}, $\alpha =1$. Given that the DM local density is very similar in all cases, the effect is small, and in Fig.~\ref{fig:all_upper_lim} we plot the case with $\rho=0.4\ \textrm{GeV/cm}^3$, taking into account stochastic effects. See Appendix~\ref{sec:app_stochasticity} for more details on the impact of including stochastic effects. The best sensitivity is found for the smallest mass considered, $m_a\approx 8\times 10^{-24}$~eV, where we set an upper limit of $g^{95\%}_{a\gamma}\approx 4\times 10^{-14}$ GeV$^{-1}$.

\subsection{Bayesian analysis}

To tackle the problem in the Bayesian framework, we first split the frequency range of interest [$10^{-9}$~Hz, $10^{-6}$~Hz] into a regular grid of 48 bins that are evenly distributed in logarithmic scale. We perform hypothesis testing and parameter estimation in each bin separately, while fixing the oscillation frequency $f$. As a result, the only varying parameter that is common for all the pulsars in Eq.~(\ref{eq:lik_fact}) is the coupling constant $g_{a\gamma}$. 

\begin{figure*}
    \centering
    \includegraphics[width=0.8\textwidth]{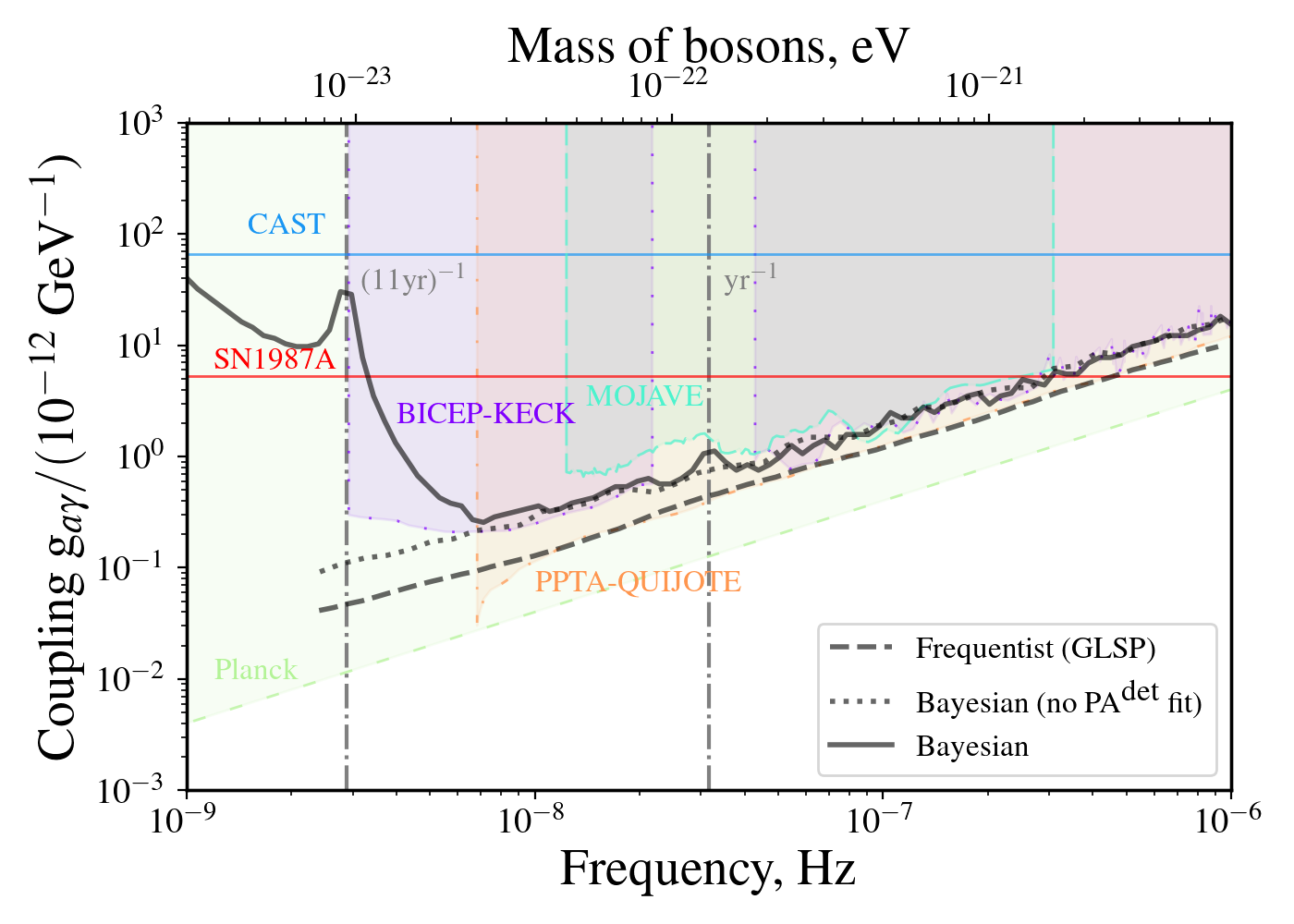}
    \caption{\label{fig:bayes_upper_lim_noG} Upper limits on the photon-ULDM coupling, $g_{a \gamma}$, obtained with the frequentist (dashed gray) and Bayesian (solid gray for the conservative approach and dotted gray for the optimistic analysis) analysis over a range of frequencies and masses. For the optimistic bayesian and frequestist analyses we cut below a minimum frequency $\nu_{\rm min}=1/T\approx 2.4\times 10^{-9}$ Hz, which corresponds to demand at least a full oscillation cycle in our data.  Other constraints obtained in previous studies are also shown: the horizontal blue line marks the start of the upper limit set by the helioscope experiment CAST~\cite{Cast2017}, likewise the red horizontal line shows the limit set by the non-observation of a gamma-ray signal from SN1987A produced by axions~\cite{Payez:2014xsa}. The area in cyan (marked with long dashes) shows the parameter space excluded by the search of the birefringence effect in AGN polarisation observations by MOJAVE VLBA~\cite{Ivanov:2018byi}, in purple (dotted) we show the area excluded by the BICEP-Keck search for the birefringence effect in CMB polarisation~\cite{BICEPKeck:2021sbt}, the orange area (dotted dashed) is excluded by the same search in polarisation data from PPTA and QUIJOTE~\cite{Castillo:2022zfl} and lastly the green area (dashed) is excluded by the non-observation of a `washout effect' in the polarisation of the CMB produced at last scattering using Planck~\cite{Fedderke:2019ajk}.}
    \label{fig:all_upper_lim}
\end{figure*}

\begin{figure}
    \centering
\includegraphics[width=\columnwidth]
{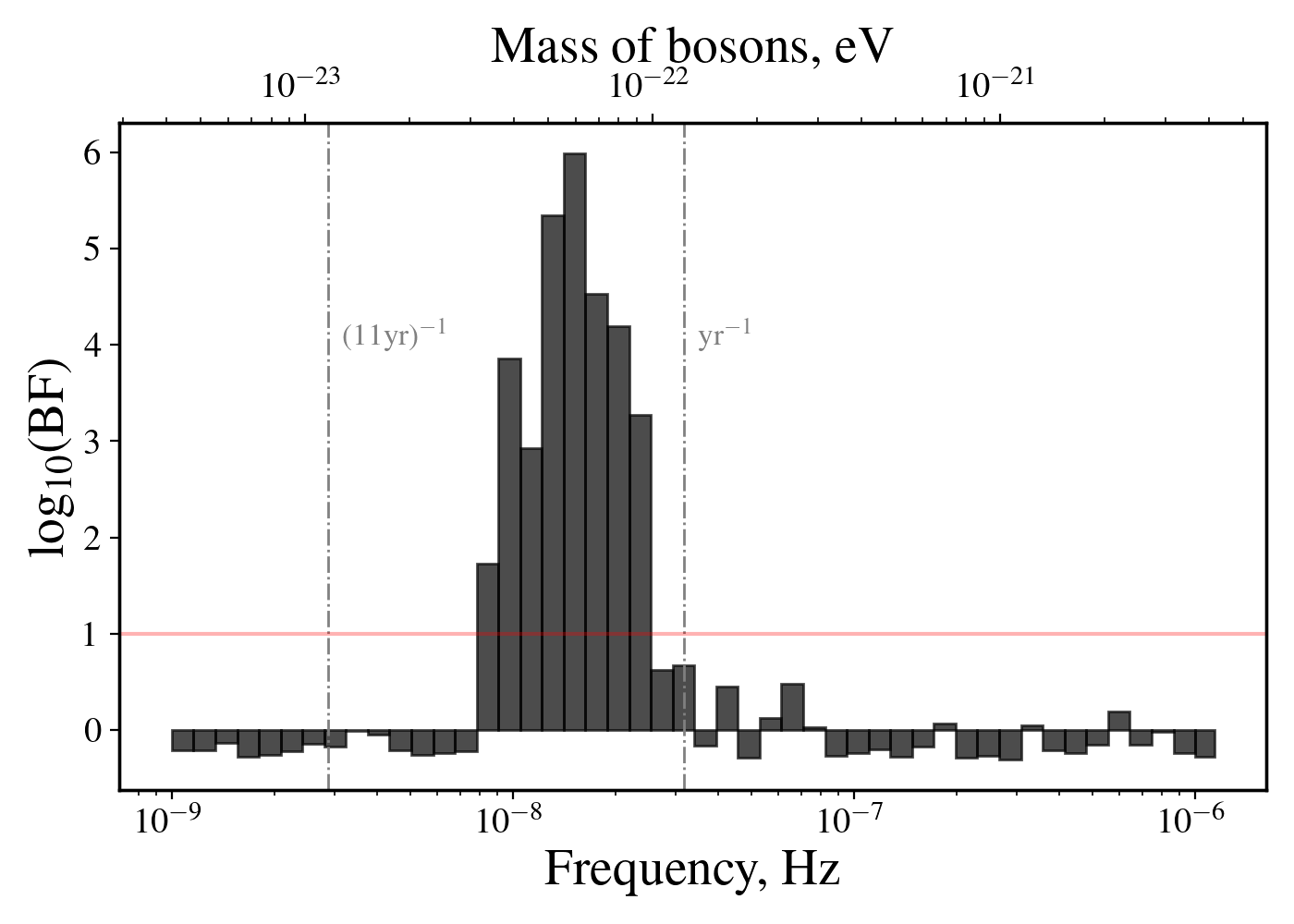}
\caption{Log$_{10}$(BF) computed using Savage-Dickey ratio from the joint posterior probabilities reconstructed from the individual posteriors distributions using the method of factorised likelihood. The red line indicates a threshold value of log$_{10}$(BF)=1.}
\label{fig:combined_bfs}
\end{figure}

\begin{figure}[h]
    \centering
\includegraphics[width=\columnwidth]
{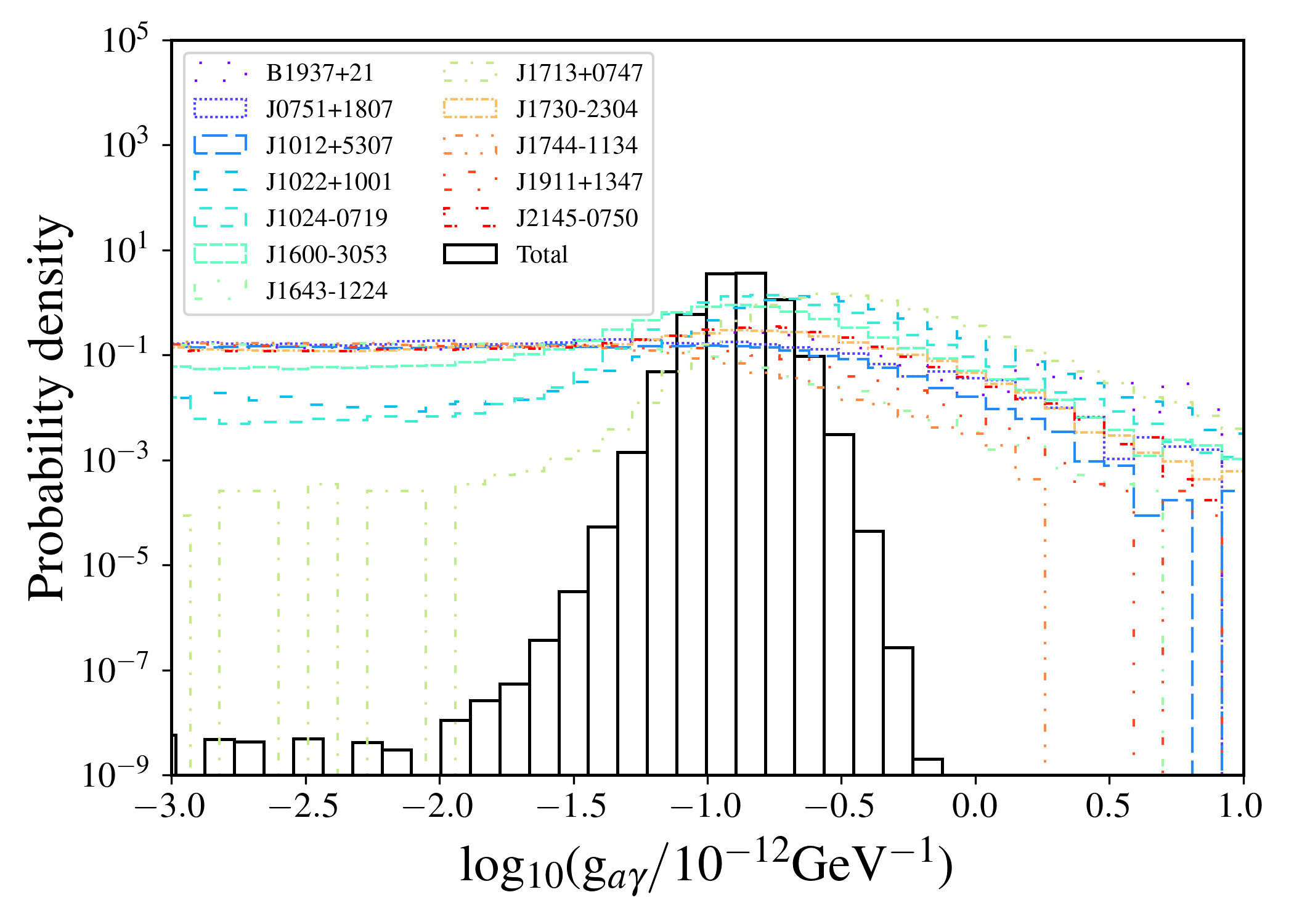}
\caption{Factorised likelihood analysis on the log$_{10}(g_{a\gamma})$ from the EPTA dataset at $f=1.4\times10^{-8}$\,Hz. Individual posterior probabilities for different pulsars are highlighted with colored (dashed and dotted dashed) histograms. The joint posterior probability is in black.}
\label{fig:fact_lik_real}
\end{figure}

The Markov chain Monte Carlo sampler ``\texttt{emcee}'' \cite{2013PASP..125..306F} is used to estimate the posterior distribution of the model parameters for each pulsar separately. The joint posterior distribution of $g_{a\gamma}$ is reconstructed by taking a product of these individual pulsar posteriors (see Eq.~(\ref{eq:lik_fact})). The obtained distributions are used to both estimate the BF in the Savage-Dickey approximation (see Eq.~(\ref{eq:SD})) and set upper limits on the coupling constant. We choose two different kinds of priors for $g_{a\gamma}$: {\it i)} uniform in linear scale for model comparison; and {\it ii)} log$_{10}$-uniform for parameter estimation. The obtained BFs are shown in Fig.~\ref{fig:combined_bfs}. For most frequency bins log$_{10}$(BF) is less than 1, indicating that no common harmonic signal was found at these particular frequencies. There are six bins between  $10^{-8}$\,Hz and $2\times 10^{-8}$\,Hz with log$_{10}$(BF) larger than 3, meaning that the hypothesis that a signal with a period of roughly 2~yrs is present in the data is supported. This frequency corresponds to a boson mass of $m_a\sim 5\times 10^{-23}$~eV. In Fig.~\ref{fig:fact_lik_real}, we show the factorised likelihood for one of the bins with $f=1.4\times10^{-8}$~Hz, clearly showing non-flat posteriors for multiple pulsars. Notably, a similar bump, but at half the frequency has been detected in \cite{epta2023, epta2023_dm, 2023ApJ...951L..50A}, when searching for continuous GWs and ULDM signatures in the latest EPTA DR2 and North American Nanohertz Observatory for Gravitational Waves datasets, respectively. Although such a signal could arise for different reasons, the most viable is the residual Faraday rotation in the terrestrial plasma which is not sufficiently well-modelled by the ionospheric model, or the effect of the Solar wind. 

\begin{figure*}
    \centering
\includegraphics[width=\columnwidth]
{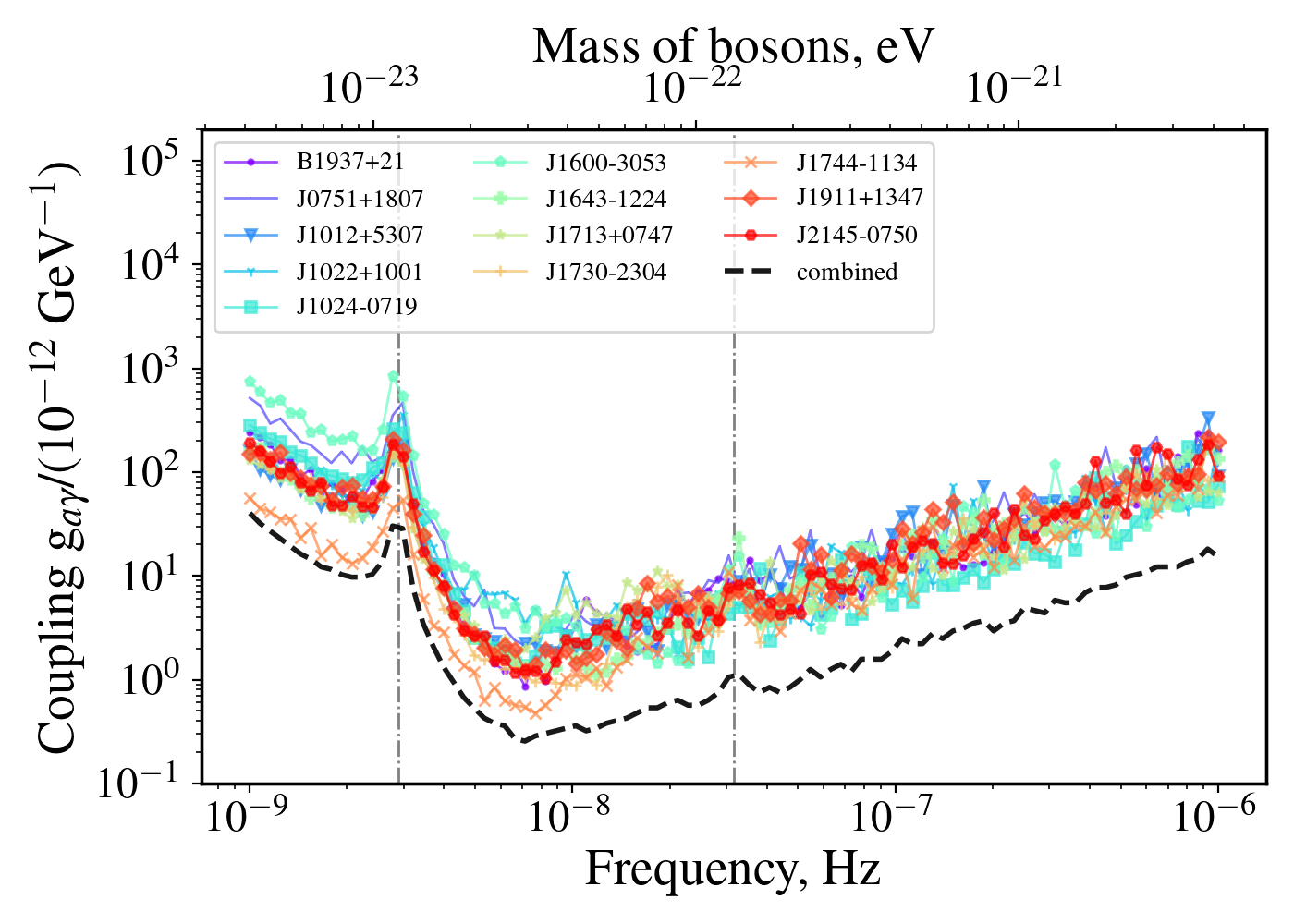}
\includegraphics[width=\columnwidth]
{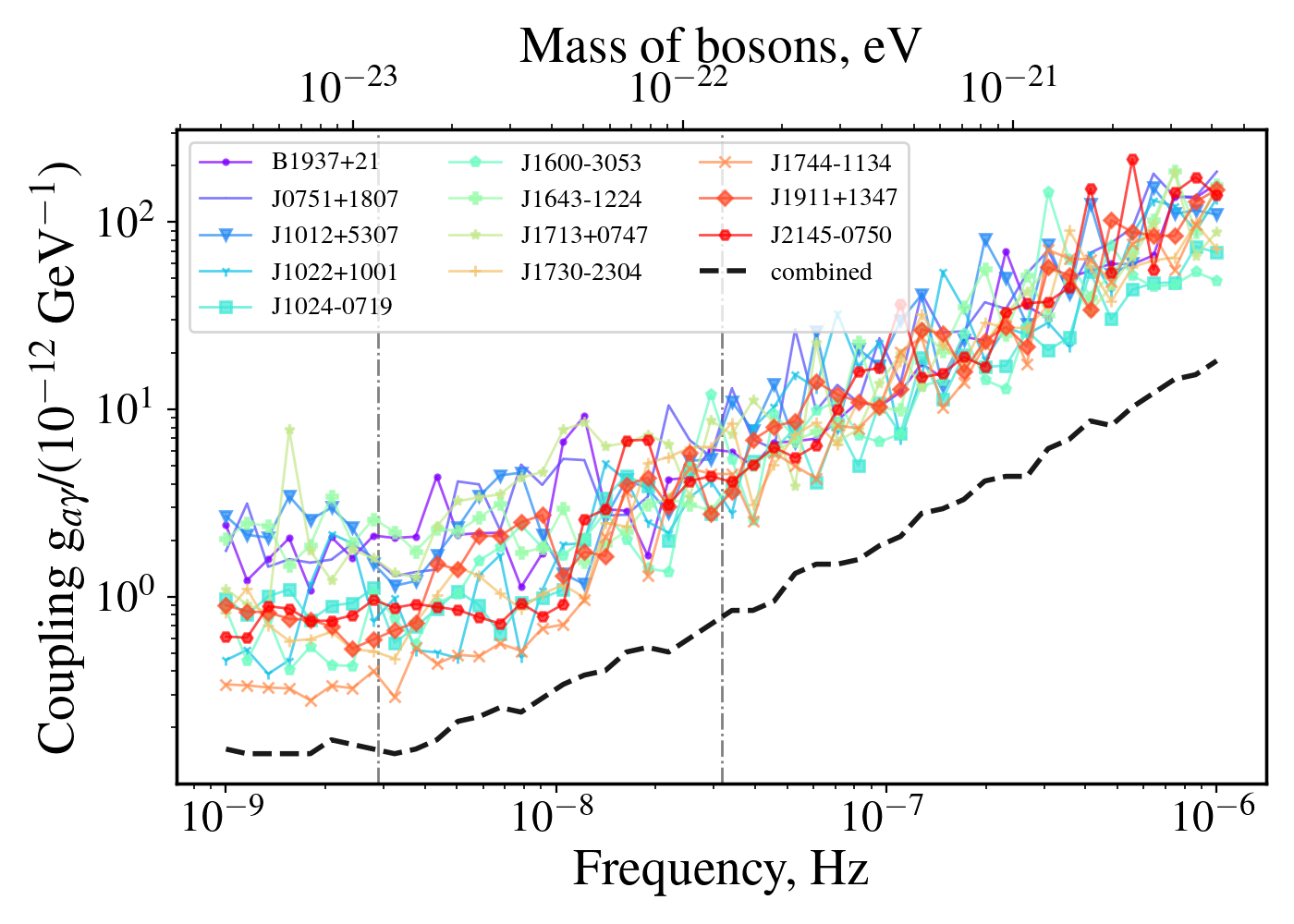}
\caption{Bayesian upper limits on the coupling constant $g_{a\gamma}$ as a function of frequency (boson mass). The different colours show the upper limits obtained for each pulsar separately, while the joint analysis is shown in black. Conservative upper limits for which the full deterministic model is taken into account are displayed on the left, while more optimistic upper bounds, for which the deterministic terms are excluded from the analysis, are shown in the right panel.}
\label{fig:bayes_upper_lim}
\end{figure*}

The upper limits on $g^{95\%}_{a\gamma}$, which are defined as 95\%-quantile of the posterior distribution, are shown in the left panel of Fig.~\ref{fig:bayes_upper_lim}. For \textit{conservative} analysis, within which a fit for nuance deterministic parameters is taken into account, the most stringent bounds are obtained at $m_a\simeq3\times10^{-23}$ eV, constraining $g_{a\gamma}$ to be below $2\times10^{-13}$ GeV$^{-1}$ with 95\% confidence. In addition, we compute \textit{optimistic} upper bounds, where we exclude a fit for deterministic systematics $\textrm{PA}^{\textrm{det}}$ from the model. The latter can be directly compared with the frequentist results. These upper limits are shown in the right panel of Fig.~\ref{fig:bayes_upper_lim} and, as expected, provide more stringent constraints at lower frequencies.

The validation of the Bayesian pipeline as well as the method of factorised likelihood is demonstrated in App.~\ref{sec:bayes_validt}, where a simulated signal is injected in the real dataset and successfully recovered.

In Fig.~\ref{fig:all_upper_lim} we show the final results of our analysis for both frequentist and Bayesian analysis for the complete twelve-pulsar dataset. Our bounds are overplotted with the results from other campaigns aimed at searching for ULDM signals in other astrophysical observables. 

\section{Conclusions}
\label{sec:conclusion}
In this paper, we analysed the polarisation data of the latest EPTA DR2 with the purpose of searching for signatures of pseudo-scalar ULDM  (axion-like particle ALP) coupled to the EM field. In order to reduce the complexity of our analysis, we focused on full Stokes observations of twelve pulsars gathered with the two largest radio telescopes in Europe, Effelsberg and Nan\c{c}ay. The presence of ULDM, interacting with the EM field with the coupling described by Eq.~\eqref{eq:Lagrangian}, results in sinusoidal variations of the PA residuals, that we searched for with both frequentist and Bayesian techniques. Another special feature of the effect is its achromaticity, i.e. it does not depend on observing frequency so that it can be effectively decoupled in the analysis from the Faraday rotation caused by the magnetised plasma.

In the frequentist framework, the search and parameter estimation have been performed using GLSP. No evidence for ULDM was found in the data, and we have obtained stringent upper limits across a wide range of ALP masses. Our best sensitivity is achieved at $m_a=8\times10^{-24}$ eV, where we get the 95\%-upper limit on the coupling constant $g^{95\%}_{a\gamma}=5\times10^{-13}$ GeV$^{-1}$.

For the Bayesian analysis, the model was extended to include additional fits for linear trends, and 11-yr and 1-yr periodic signals that may be present in the data due to polarisation calibration issues and residual ionospheric systematics. We have performed a frequency-resolved analysis by splitting the frequency band of interest into multiple segments and performing search and parameter estimation independently in each of them. We find strong evidence for a monochromatic signal in 6 bins in the frequency range between $10^{-8}$\,Hz and $2\times 10^{-8}$\,Hz. The corresponding coupling constant of the detected signal is $\textrm{log}_{10}(g_{a\gamma}/\textrm{GeV}^{-1})=-13.0\pm0.86$, which is in marginal contradiction with Planck and PPTA-QUIJOTE constraints (see Fig.~\ref{fig:all_upper_lim}). The nature of this signal is still unclear, but it is most likely related to poorly modelled Faraday rotation in the terrestrial plasma.

The most stringent constraints on the coupling constant in the Bayesian framework are obtained at $m_a=3\times 10^{-23}$~eV, where  $g^{95\%}_{a\gamma}=2\times 10^{-13}$ GeV$^{-1}$. Additionally, we have calculated Bayesian upper limits for a simplified model (excluding deterministic systematics), which are in good agreement (within a factor of 2) with the frequentist results. 

There are a number of reasons that can cause the difference between the two types of analysis. As was discussed earlier, in contrast to frequentist analysis, Bayesian models include white noise modifiers, EFAC and EQUAD, that alter the uncertainties of the obtained model parameters, in particular, upper limits on $g_{a\gamma}$. Moreover, the GLSP analysis is constructed under the assumption that the stochastic processes are white. As it was shown in \cite{2011MNRAS.418..561C}, the presence of coloured noises can introduce severe errors in the GLSP parameter estimation and detection statistics. In case of the Bayesian framework, correlated systematics have been partially accounted for by introducing long-term deterministic trends in the model (as a proxy for the "red noise").

Some interesting technical issues relating to the search methods and the nature of the dataset have also emerged from this analysis. They can be summarised as follows:
\begin{itemize}
\item The subtraction of an averaged PA template from individual PA profiles does not result in white Gaussian residuals, leading to a significant underestimation of the uncertainties on the resulting epoch-averaged PAs. A non-negligible change in the PA profile shapes is the main reason for this behaviour. The nature of these variations is uncertain but is most likely related to temporal variations in the IISM scattering, which can cause significant distortion of the PA swing (see \cite{k2009}). Imperfect calibration, e.g. caused by the residual leakages between two orthogonal receivers, could provide another possible explanation. Similar underestimated uncertainties of the PA residuals and PA profile temporal variations were obtained with other radio instruments reported in \cite{2011Ap&SS.335..485Y, 2013MNRAS.430..416O, 2024arXiv240613463D}. The characteristics of the systematics that appear in the phase-resolved PA and possible ways to reduce them will be investigated in future studies. 
\item The Faraday rotation in the terrestrial ionospheric plasma causes a dynamical shift of the PA and can also be responsible for the distortion of the PA swing described above. Despite of being chromatic, the Faraday rotation can still be responsible for the emergence of ULDM-like signals and subsequent false detection. For instance, we suspect that the excess of power that was detected in the current analysis may have been the result of the ionosphere not being modelled correctly. The ionospheric contribution can be accounted for with the external models of the ionosphere, as done in the current analysis. The shortcoming of this scheme is that none of the existing models can accurately describe the complex physics of the magneto-ionic content of the Earth's atmosphere. This problem can be overcome by performing a similar study at lower observing frequencies, in order to break the degeneracy between the chromatic Faraday rotation and the achromatic ULDM signal. In this case, the data can be \textit{self-calibrated} for varying Faraday rotation (both, in the ionospheric and IISM) without the use of third-party models.
\item For both the frequentist and Bayesian analysis, we have neglected any possible inter-pulsar correlations caused by the ULDM signal and effectively looked for an incoherent monochromatic signal in different pulsars. It is important to note that the astrophysical signal of interest may be partially monopolar-correlated. One possible way of including cross-correlations in the analysis was considered in \cite{PhysRevLett.130.121401}. Determining whether the detected signal exhibits any angular correlation can be performed in post-processing (similar to the optimal statistics analysis of the GW background \cite{2015PhRvD..91d4048C}) and is going to be addressed in subsequent work. 
\item Our upper limits on the coupling constant can be ameliorated by selecting pulsars located in regions of high DM density. Millisecond pulsars that are located close to the Galactic centre and are known to have stable profiles will provide a significant increase in sensitivity towards ULDM signals. In particular, long-term observations of the highly linearly polarised PSR J1744$-$2946, which was recently discovered in a Galactic centre filament, would be an excellent candidate to test the ULDM hypothesis \cite{ldj+2024}.
\end{itemize}

The EPTA collaboration is an ongoing effort to detect GWs through performing regular observations of MSPs over several decades. The same data can be used to study a wide range of astrophysical problems beyond the detection of GWs. As it was shown in the current paper, searches for ultra-light DM candidates is one of the good examples. Here, we restrict our analysis to the data from Effelsberg and Nan\c{c}ay radio telescopes only. In the future, we plan to extend our analysis to include the data from three more observatories, namely JBO, SRT, and WSRT, as well as the LOFAR interferometer. LOFAR has recently been incorporated into the EPTA collaboration, and will provide low-frequency support to the campaign. Upcoming data releases with longer baselines and updated instrumentation are expected to further increase our sensitivity to ALPs, maximising the potential for a detection of ULDM.
\section{Data Availability Statement}
The data that support the findings of this article are openly available \cite{porayko_2024_14005957}, embargo periods may apply. Additional data products
used in the current manuscript can be provided upon reasonable request.
\nocite{*}

\begin{acknowledgements}
The European Pulsar Timing Array (EPTA) is a collaboration between
European and partner institutes, namely ASTRON (NL), INAF/Osservatorio
di Cagliari (IT), Max-Planck-Institut f\"{u}r Radioastronomie (GER),
Nan\c{c}ay/Paris Observatory (FRA), the University of Manchester (UK),
the University of Birmingham (UK), the University of East Anglia (UK),
the University of Bielefeld (GER), the University of Paris (FRA), the
University of Milan-Bicocca (IT) and Peking University (CHN), with the
aim to provide high precision pulsar timing to work towards the direct
detection of low-frequency gravitational waves. An Advanced Grant of
the European Research Council to implement the Large European Array
for Pulsars (LEAP) has also provided funding. The EPTA is part of the
International Pulsar Timing Array (IPTA); we would like to thank our
IPTA colleagues for their help with this paper.

Part of this work is based on observations with the 100-m telescope of
the Max-Planck-Institut f\"{u}r Radioastronomie (MPIfR) at Effelsberg
in Germany. Pulsar research at the Jodrell Bank Centre for
Astrophysics and the observations using the Lovell Telescope are
supported by a Consolidated Grant (ST/T000414/1) from the UK's Science
and Technology Facilities Council (STFC). ICN is also supported by the
STFC doctoral training grant ST/T506291/1. The Nan{\c c}ay radio
Observatory is operated by the Paris Observatory, associated with the
French Centre National de la Recherche Scientifique (CNRS), and
partially supported by the Region Centre in France. We acknowledge
financial support from ``Programme National de Cosmologie and
Galaxies'' (PNCG), and ``Programme National Hautes Energies'' (PNHE)
funded by CNRS/INSU-IN2P3-INP, CEA and CNES, France. We acknowledge
financial support from Agence Nationale de la Recherche
(ANR-18-CE31-0015), France. The Westerbork Synthesis Radio Telescope
is operated by the Netherlands Institute for Radio Astronomy (ASTRON)
with support from the Netherlands Foundation for Scientific Research
(NWO). The Sardinia Radio Telescope (SRT) is funded by the Department
of University and Research (MIUR), the Italian Space Agency (ASI), and
the Autonomous Region of Sardinia (RAS) and is operated as a National
Facility by the National Institute for Astrophysics (INAF).

The work is supported by the National SKA programme of China
(2020SKA0120100), Max-Planck Partner Group, NSFC 11690024, CAS
Cultivation Project for FAST Scientific. This work is also supported
as part of the ``LEGACY'' MPG-CAS collaboration on low-frequency
gravitational wave astronomy. JA acknowledges support from the
European Commission (Grant Agreement number: 101094354), the Stavros
Niarchos Foundation (SNF) and the Hellenic Foundation for Research and
Innovation (H.F.R.I.) under the 2nd Call of ``Science and Society --
Action Always strive for excellence -- ``Theodoros Papazoglou''
(Project Number: 01431). AF, AS, DIV, GS, MB acknowledge
financial support provided under the European Union's H2020 ERC
Consolidator Grant ``Binary Massive Black Hole Astrophysics'' (B
Massive, Grant Agreement: 818691). AC and AP acknowledge financial 
support from the European Research Council (ERC) starting grant 'GIGA' 
(grant agreement number: 101116134). GD, RK and MK acknowledge support
from European Research Council (ERC) Synergy Grant ``BlackHoleCam''
Grant Agreement Number 610058 and ERC Advanced Grant ``LEAP'' Grant
Agreement Number 337062. JWMK is a CITA Postdoctoral Fellow: This work
was supported by the Natural Sciences and Engineering Research Council
of Canada (NSERC), (funding reference CITA 490888-16). KP acknowledges support from the State project ``Science'' by the Ministry of Science and Higher Education of Russia under the contract 075-15-2024-541. AV acknowledges the support of the Royal Society and Wolfson Foundation, and the UK Science and Technology Facilities Council (STFC). JPWV acknowledges
support by the Deutsche Forschungsgemeinschaft (DFG) through the
Heisenberg programme (Project No. 433075039) and by the NSF through
AccelNet award \#2114721. NKP funded by the Deutsche
Forschungsgemeinschaft (DFG, German Research Foundation) --
Projektnummer PO 2758/1--1, through the Walter--Benjamin
programme. ASamajdar thanks the Alexander von Humboldt foundation in
Germany for a Humboldt fellowship for postdoctoral researchers. AP, DP
and MB acknowledge support from the research grant “iPeska”
(P.I. Andrea Possenti) funded under the INAF national call
Prin-SKA/CTA approved with the Presidential Decree 70/2016
(Italy). RNC acknowledges financial support from the Special Account
for Research Funds of the Hellenic Open University (ELKE-HOU) under
the research programme ``GRAVPUL'' (grant agreement 319/10-10-2022).
EvdW, CGB and GHJ acknowledge support from the Dutch National Science
Agenda, NWA Startimpuls – 400.17.608.

The work of JTC is supported by the Ministerio de Ciencia e Innovaci\'on under FPI contract PRE2019-089992 of the SEV-2015-0548 grant, by the project “Theoretical Astroparticle Physics (TAsP)” funded by the INFN and from the research grant “Addressing systematic uncertainties in searches for dark matter No. 2022F2843L” funded by MIUR. JMC and JTC acknowledge support from the MICINN through the grant ``DarkMaps'' PID2022-142142NB-I00. D. Blas acknowledges the support from the Departament de Recerca i Universitats from Generalitat de Catalunya to the Grup de Recerca 00649 (Codi: 2021 SGR 00649). EB acknowledges support from the European Union’s Horizon Europe programme under the Marie Sklodowska Curie grant agreement No 101105915 (TESIFA), the European Consortium for Astroparticle Theory in the form of an Exchange Travel Grant, the European Union’s Horizon 2020 Programme under the AHEAD2020 project (grant agreement n. 871158), and the European Research Council (ERC) under the European Union’s Horizon 2020 research and innovation program ERC2018-CoG undergrant agreement N. 818691 (B Massive).
The research leading to these results has received funding from the Spanish Ministry of Science and Innovation (PID2020-115845GB-I00/AEI/10.13039/501100011033).
IFAE is partially funded by the CERCA program of the Generalitat de Catalunya. Finally JMC and DB acknowledge funding received from the European Union through the grant ``UNDARK'' of the Widening participation and spreading excellence programme (project number 101159929).

We made extensive use of publicly available \texttt{numpy}~\cite{2020Natur.585..357H}, \texttt{scipy}~\cite{2020NatMe..17..261V}, \texttt{matplotlib}~\cite{2007CSE.....9...90H}, \texttt{emcee}~\cite{2013PASP..125..306F} python libraries as well as \texttt{PSRCHIVE} pulsar software~\cite{Hotan+2004}.
\end{acknowledgements}
\appendix

\section{Influence of coherence patch size}
\label{sec:app_stochasticity}

\begin{figure}
    \centering
\includegraphics[width=\columnwidth]
{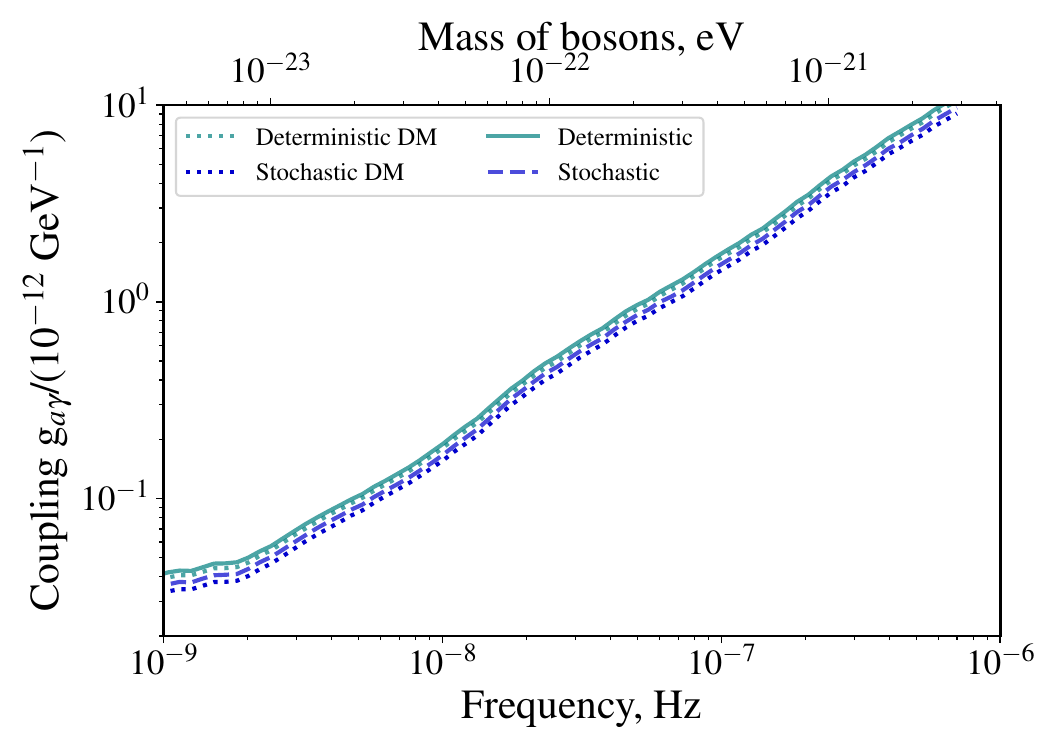}
\caption{Frequentist upper limits on the coupling constant $g_{a\gamma}$ produced in four different regimes of `coherence'. See text on details of these modes. }
\label{fig:stoch}
\end{figure}

As stated in the main text, the connection between the axionic field amplitude and the dark matter density takes a stochastic nature, see Eq.~\eqref{eq:alpha}, that follows a Rayleigh distribution in each coherence domain~\cite{Foster:2017hbq,Centers:2019dyn}. With the coherence length
 \begin{equation}
 \label{eq:coherenced}
     l_{c} =\left(m_{a}\sigma\right)^{-1}\simeq 65  \left(\frac{m_{a}}{10^{-22} \text{ eV}}\right)^{-1}\left(\frac{\sigma}{10^{-3}}\right)^{-1} {\rm pc},
 \end{equation}
 we can determine whether two or more pulsars are in the same coherence patch for a given ULDM mass. For masses below $\sim 5\times 10^{-23}$~eV, there are several pulsars that lie in the same coherence patch. Then, to calculate the DM local density in each source and the Earth we have used a Navarro-Frenk-White (NFW) \cite{1996ApJ...462..563N, 1997ApJ...490..493N} profile with the parameters $r_s = 14.48$ kpc and $\rho_s=0.566\  \rm GeV/cm^3$~\cite{Cirelli:2024ssz}. To see the effect of assuming a common DM density, $\rho_{\rm DM}\approx 0.4\ \rm GeV/cm^3$, and a common coherence patch we performed the global frequentist analysis in four possible cases, as shown in Fig.~\ref{fig:stoch}. We labelled the analysis as \textit{deterministic} when we neglect stochastic effects, i.e. $\alpha_0=\alpha_p=1$, and \textit{DM} when the local DM density assuming NFW profile has been implemented for each pulsar. We can see that the result does not change significantly. For the main result presented in Fig.~\ref{fig:all_upper_lim}, we choose the stochastic case with equal DM density.

\section{Validation of the Bayesian search pipeline}
\label{sec:bayes_validt}
\begin{figure*}
    \centering
\includegraphics[width=0.8\textwidth]
{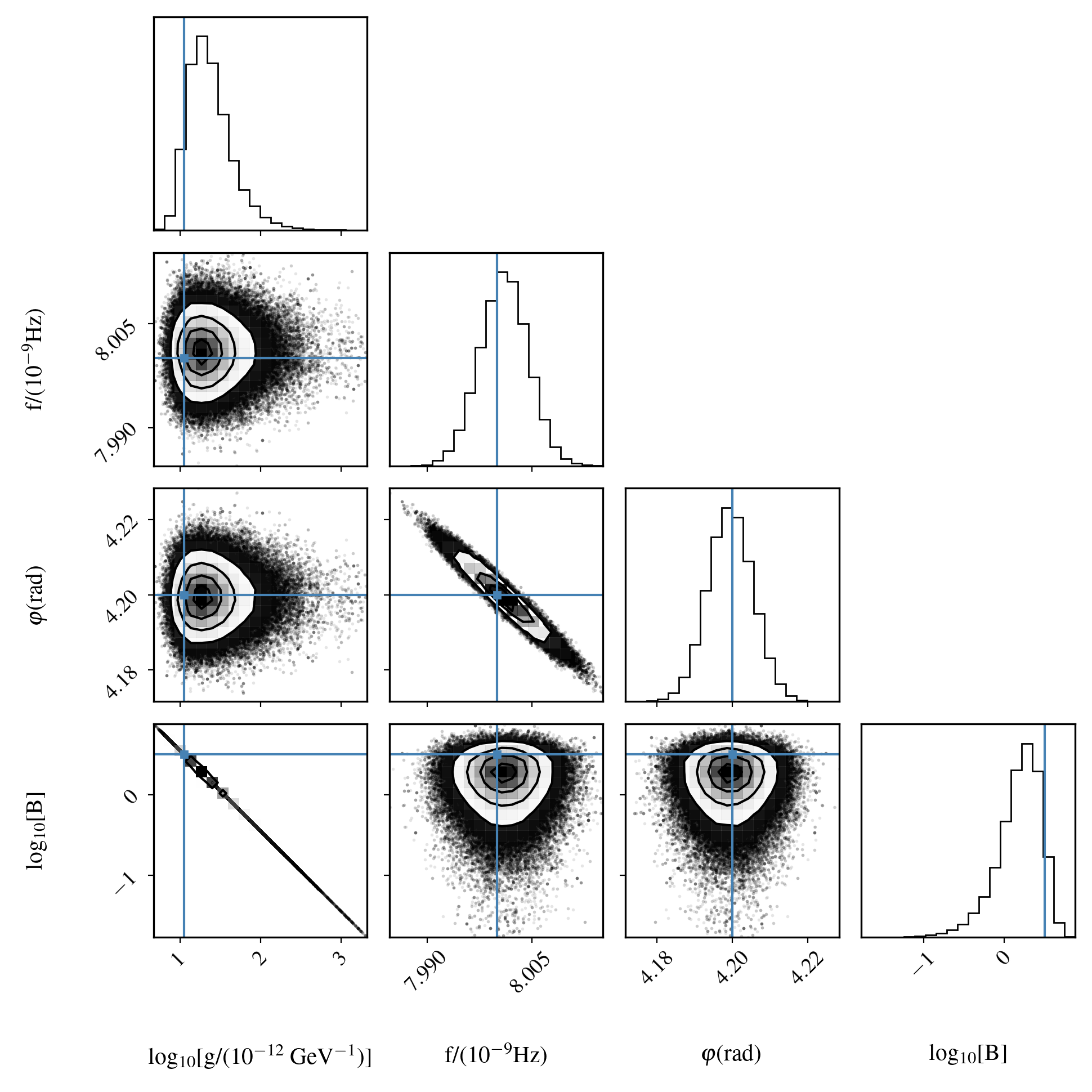}
\caption{Posterior distribution of astrophysical model parameters which are summarised in Table~\ref{tab:bayes_params}. The simulated signal has been injected into the real data of PSR~J1744-1134. The injected values are shown with blue solid lines and the contours are 1-, 2- and 3-$\sigma$ intervals.}
\label{fig:recover1}
\end{figure*}

\begin{figure}
    \centering
\includegraphics[width=\columnwidth]
{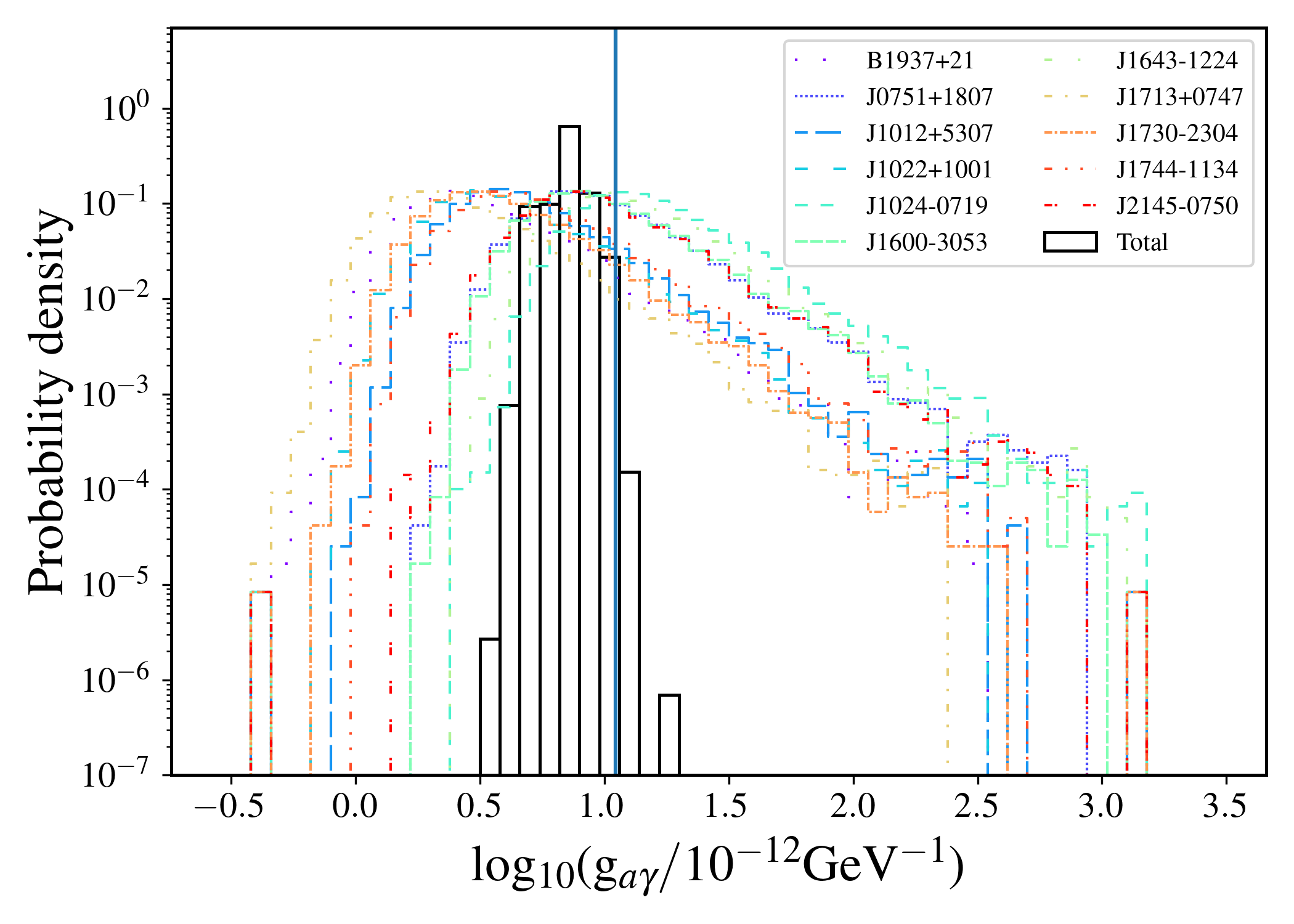}
\caption{Factorised likelihood analysis on $g_{a\gamma}$. The posterior distributions of individual pulsars are shown in colour (dashed and dotted dashed), while the joint posterior is in black. The injected value is highlighted in blue. One should note that the plot shows the marginalised distribution of a 2-D ($g_{a\gamma}$, $m_a$) posterior.}
\label{fig:fact_lik}
\end{figure}

To validate our methods and check the robustness of our Bayesian schemes, we have performed an injection-recovery test. The simulated ULDM signal was injected into the real pulsar polarimetry data used in the current study. In order to recover the signal, we run a fully Bayesian analysis as described in Sec~\ref{sec:bayes_analysis}. The posterior distribution of astrophysical parameters for PSR~J1744$-$1134 are shown in Fig.~\ref{fig:recover1}. The joint posterior distribution was constructed by multiplying individual 2D-posterior distributions ($g_{a\gamma}, f$) of 12 pulsars. The factorised distribution of $g_{a\gamma}$ marginalised over $f$ is shown in Fig.~\ref{fig:fact_lik}, demonstrating the successful recovery of the injected value.

\section{Observation of the residual systematics in the current analysis}
\label{sec:iono_sys}

\begin{figure*}
\centering
\includegraphics[width=\columnwidth]
{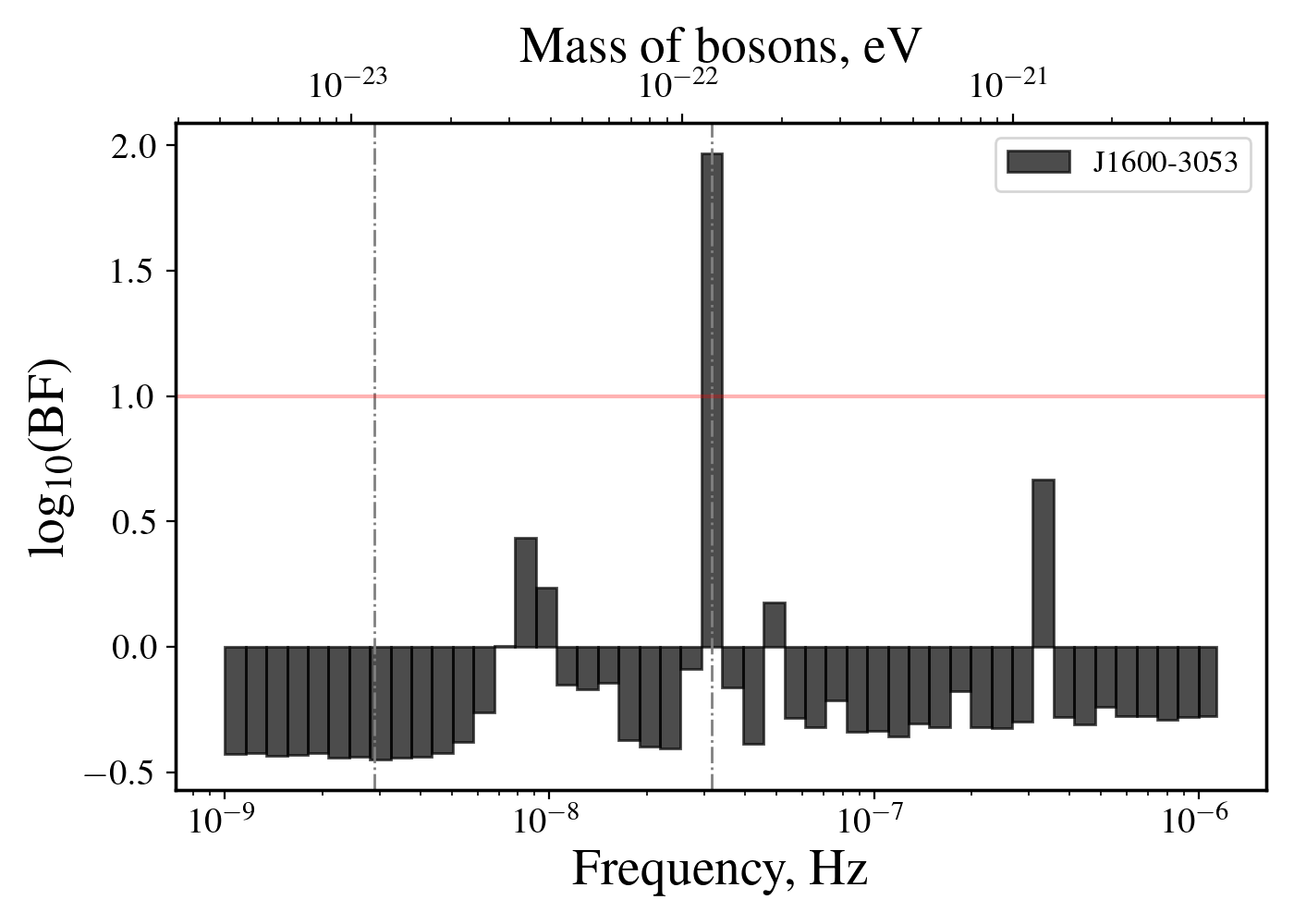}
\includegraphics[width=\columnwidth]
{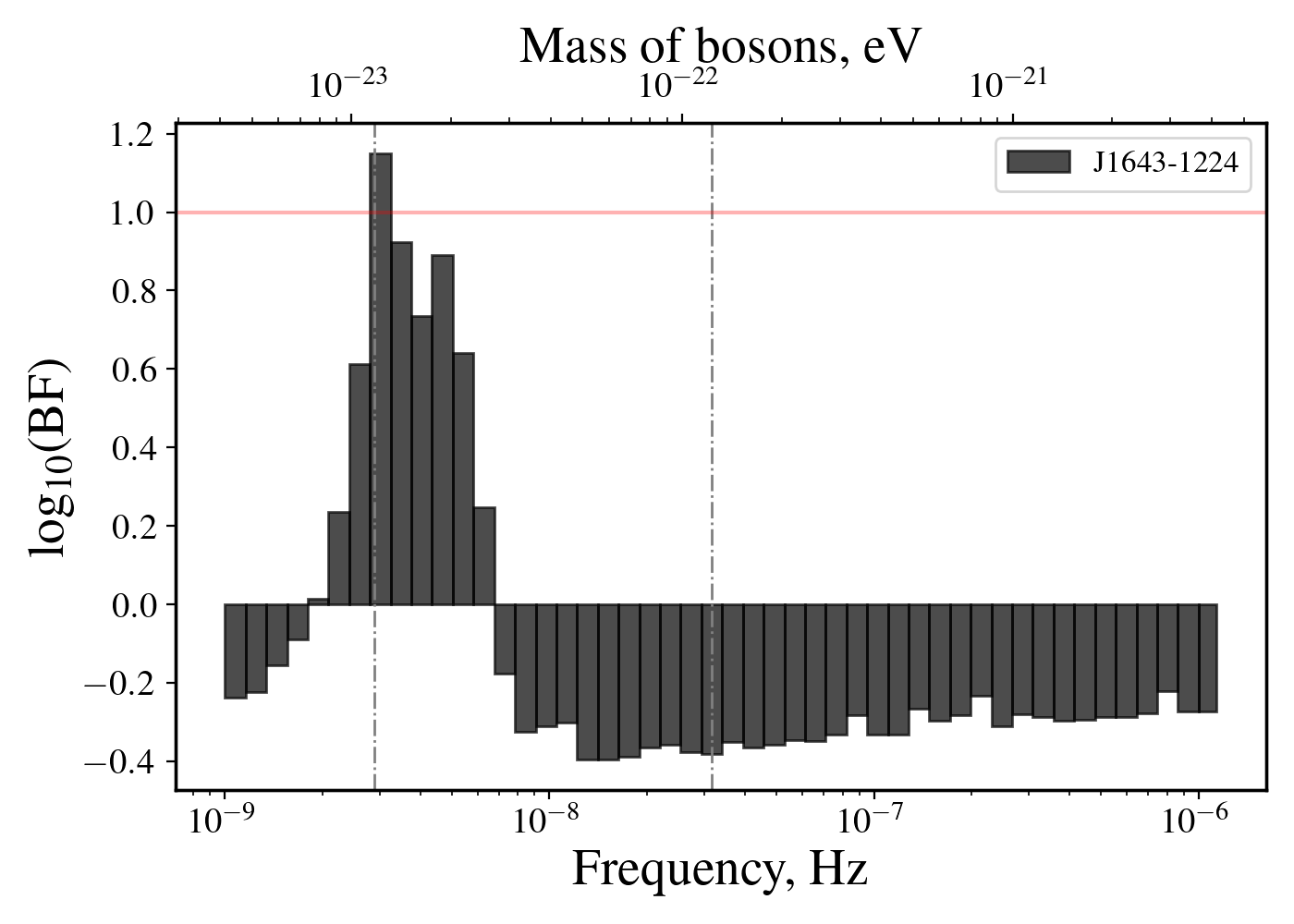}
\caption{Log$_{10}$(BF) computed using the Savage-Dickey ratio from the individual posterior distributions of PSR~J1600$-$3053 (on the left) and PSR~J1643$-$1224 (on the right). The red line indicates a threshold value of log$_{10}$(BF)=1. The calculated BFs exhibit peaks at 1-yr and 11-yr.}
\label{fig:sys_bfs}
\end{figure*}

\begin{figure*}
\centering
\includegraphics[width=\columnwidth]
{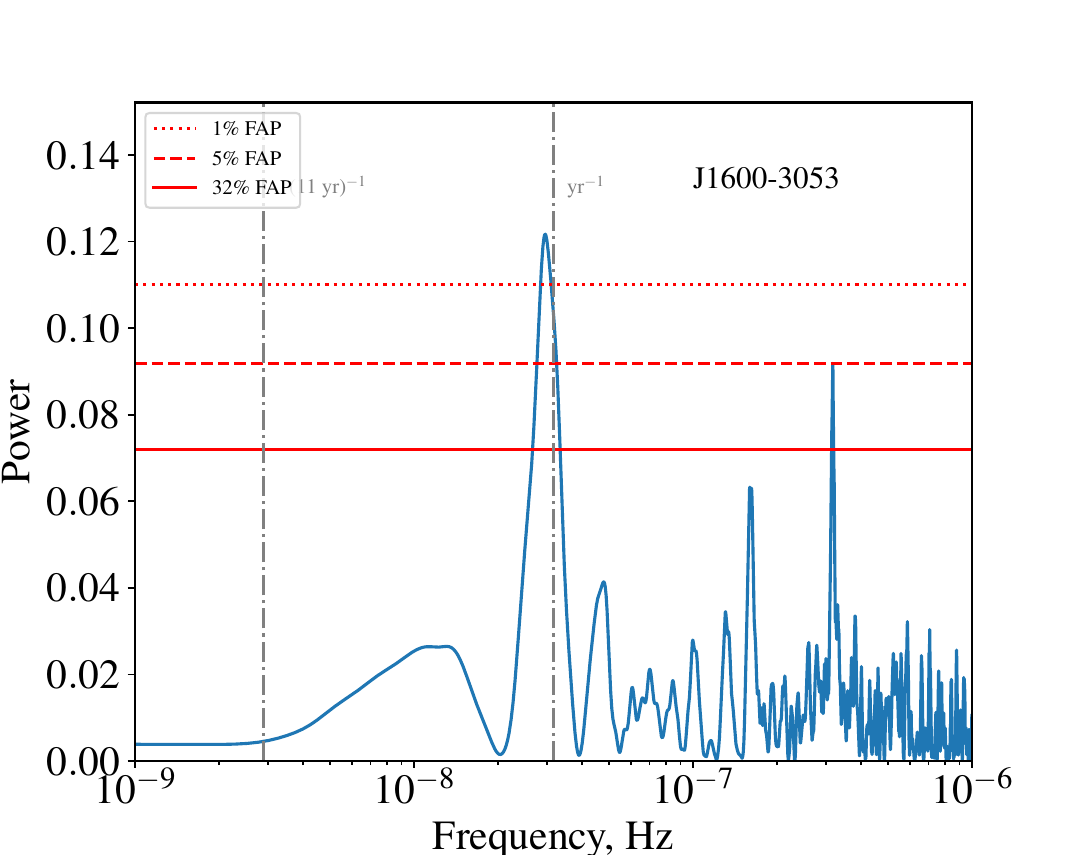}
\includegraphics[width=\columnwidth]
{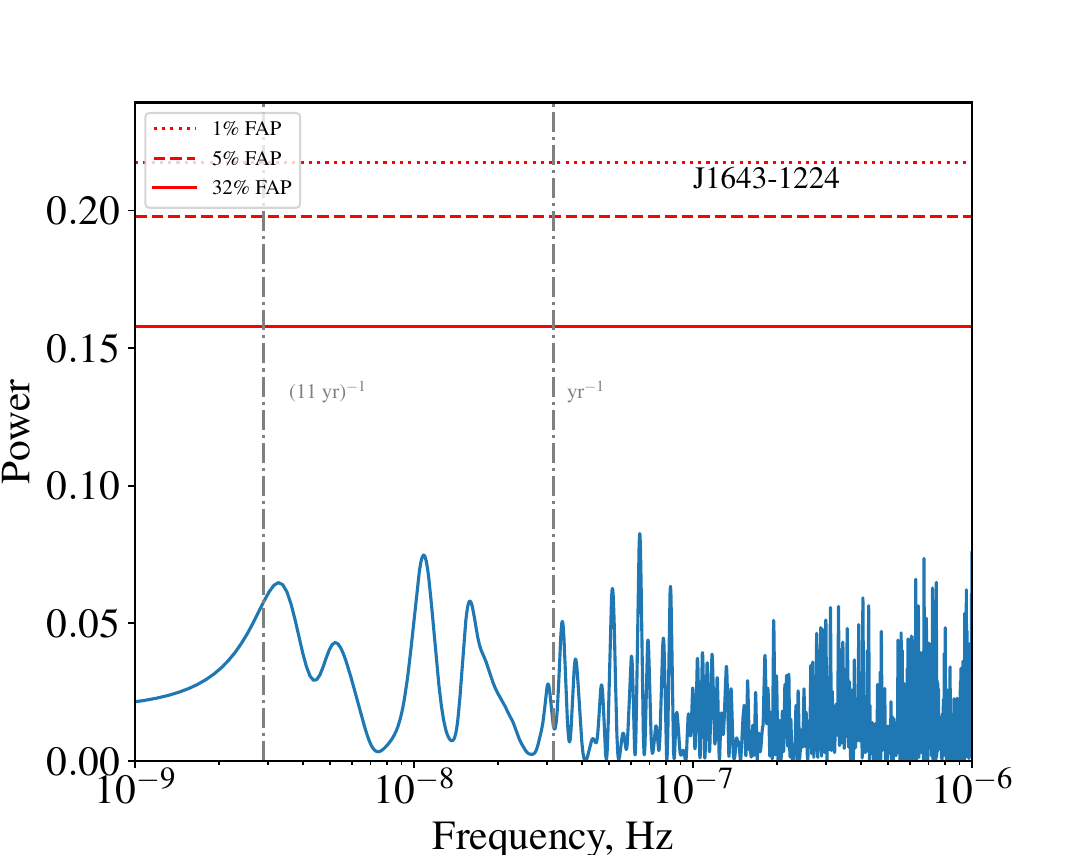}
\caption{Generalised Lomb-Scargle periodogram and the corresponding FAP at 1\%, 5\% and 32\% of PSR~J1600$-$3053 (on the left) and PSR~J1643$-$1224 (on the right). We observe a peak at 1-yr in PSR~J1600$-$3053 above the FAP level.}
\label{fig:Periodogram}
\end{figure*}

After ionospheric corrections, see Sec.~\ref{sec:iono_correction}, some spurious signals from the modelling might still be left. In this section, we highlight some of these signatures that were detected in our dataset within both frequentist and Bayesian frameworks. For such a demonstration, we removed the $\textbf{PA}^{\textrm{det}}$ from the Bayesian model. For example, in PSR~J1600$-$3053 we found a signal compatible with one year periodicity (see the left side of Figs.~\ref{fig:sys_bfs} and \ref{fig:Periodogram}). For the case of PSR~J1643$-$1224, in the Bayesian analysis a signal with a frequency of (11 yr)$^{-1}$ emerged. However, in the frequentist analysis, while there is a peak with relatively high power at that frequency, it is not statistically significant, i.e. it is not above the FAP level (see the right side of Figs.~\ref{fig:sys_bfs} and \ref{fig:Periodogram}). Both signals are well-known residual quasi-periodicities after the subtraction of the SLM ionospheric model corresponding to a seasonal effect ($\sim$ 1 year) and to the solar cycle ($\sim$ 11 years).

\bibliography{apssamp}

\end{document}